\documentclass[prc,aps,floatfix,showpacs,twocolumn]{revtex4}
\usepackage{dcolumn}
\usepackage{bm}
\usepackage{color}
\usepackage{amssymb}
\usepackage{amsmath}
\usepackage{graphicx}
\usepackage{pstricks}
\begin{document}

\title{The two-nucleon electromagnetic charge operator in chiral effective field theory
($\chi$EFT) up to one loop}
\author{S.\ Pastore$^{\,{\rm a}}$, L.\ Girlanda$^{\,{\rm b,c}}$,  R.\ Schiavilla$^{\,{\rm d,e}}$, and
M.\ Viviani$^{\,{\rm f}}$}

\affiliation{
$^{\rm a}$\mbox{Physics Division, Argonne National Laboratory, IL 60439, USA}\\
$^{\,{\rm b}}$\mbox{Department of Physics, Universit\`a del Salento, I-73100 Lecce, Italy}\\
$^{\,{\rm c}}$\mbox{INFN Sezione di Lecce, I-73100 Lecce, Italy}\\
$^{\rm d}$\mbox{Department of Physics, Old Dominion University, Norfolk, VA 23529, USA}\\
$^{\rm e}$\mbox{Jefferson Lab, Newport News, VA 23606, USA}\\
$^{\,{\rm f}}$\mbox{INFN Sezione di Pisa, I-56127 Pisa, Italy}\\
}
\date{\today}

\begin{abstract}
The electromagnetic charge operator in a two-nucleon system is derived in
chiral effective field theory ($\chi$EFT) up to order $e\, Q$ (or N4LO), where $Q$ denotes
the low-momentum scale and $e$ is the electric charge.  The specific form
of the N3LO and N4LO corrections from, respectively, one-pion-exchange and
two-pion-exchange depends on the off-the-energy-shell prescriptions adopted
for the non-static terms in the corresponding potentials.  We show
that different prescriptions lead to unitarily equivalent potentials and accompanying
charge operators.  Thus, provided a consistent set is adopted, predictions
for physical observables will remain unaffected by the non-uniqueness associated with
these off-the-energy-shell effects.
\end{abstract}


\maketitle

\section{Introduction}
\label{sec:intro}
The chiral symmetry exhibited by quantum chromodynamics (QCD) dictates that
the pion couples to baryons, such as nucleons and $\Delta$-isobars,
by powers of its momentum $Q$.  As a consequence, the Lagrangian
describing these interactions can be expanded in powers of $Q/\Lambda_\chi$,
where $\Lambda_\chi \sim 1$ GeV specifies the chiral-symmetry breaking
scale.  Thus, classes of Lagrangians emerge, each characterized
by a given power of $Q/\Lambda_\chi$ and each involving a certain number of
unknown coefficients, so called low-energy constants, which are then
determined by fits to experimental data (see, for example, the
review papers~\cite{Bedaque02} and~\cite{Epelbaum09}, and
references therein).  This approach, known as chiral effective field
theory ($\chi$EFT), can be justifiably argued to have put
low-energy nuclear physics on a more fundamental basis by providing,
on the one hand, a direct connection between the symmetries of
QCD---in particular, chiral symmetry---and the strong and
electroweak interactions in nuclei, and, on the other hand, a practical
calculational scheme susceptible, in principle, of systematic improvement.

The model for the nuclear electromagnetic current---the space-like part
of the four current---in $\chi$EFT up to one loop was derived
originally by Park {\it et al.}~\cite{Park96}, using covariant perturbation
theory.  In the last couple of years, two independent derivations, based
on time-ordered perturbation theory (TOPT), have appeared in the literature, one
by the present authors~\cite{Pastore09} and the other by K\"olling
{\it et al.}~\cite{Koelling09}.  There are technical differences in the
implementation of TOPT, which relate to the treatment of reducible diagrams
and are documented in considerable detail in the above papers.
However, the resulting expressions in Refs.~\cite{Pastore09} and~\cite{Koelling09}
for the two-pion-exchange currents (the only ones considered by the authors
of Ref.~\cite{Koelling09}) are in agreement with each other, but differ
from those of Ref.~\cite{Park96}, in particular in the isospin structure
of the $M1$ operator associated with the one-loop corrections---see
Pastore {\it et al.} (2009)~\cite{Pastore09} for a comparison
and analysis of these differences.

K\"olling {\it et al.}~also provided the first treatment
of loop corrections to the nuclear charge operator---the time-like
part of the four current---associated with two-pion exchange
mechanisms.  Of course, there had been earlier studies of
the two-nucleon charge operator, notably those of
Refs.~\cite{Walzl01,Phillips03,Phillips07}, but they had
been limited to its isoscalar component, and therefore
had only retained tree-level corrections (the two-pion-exchange
loop contributions are isovector).  These earlier studies also
included predictions for the charge and quadrupole
form factors of the deuteron, which were in reasonable agreement
with data obtained from measurements of elastic
electron-deuteron scattering cross sections at low-momentum transfer.

The primary objective of the present work is to extend the formalism
developed in Refs.~\cite{Pastore09,Pastore08} to derive systematically the
two-nucleon electromagnetic charge operator in $\chi$EFT, including
up to one loop corrections.  As we shall see below,
this is not a straight-forward task, since the derivation of such an operator
necessarily entails the study of non-static contributions to the
one-pion-exchange (OPE) and two-pion-exchange (TPE) potentials.
In the OPE sector, this inter-connection between non-static
contributions and the charge operator was investigated
long ago by Friar~\cite{Friar77} in the context of a
Foldy-Wouthuysen reduction procedure and a time-dependent
perturbation theory, which consistently retained corrections
up to order $(v/c)^2$.  In particular, Friar showed that i)
the charge operators so derived depend on the specific, but arbitrary,
off-the-energy extension---{\it i.e.}, on the corrections beyond the static
limit, such as those induced by retardation effects---adopted
for the OPE potential, and that ii) these different operators
(and corresponding OPE potentials) are related to each other
by a unitary transformation and, therefore, their intrinsic lack
of uniqueness has no consequence on the predictions for
physical observables.

In this paper, we examine these issues from a $\chi$EFT perspective.
We are interested in constructing
the charge operators up to one loop and, hence, need to include non-static
corrections not only in the OPE, but also in the TPE potential.
We show that the resulting operators, while not unique because
of the off-the-energy-shell ambiguity referred to above, are nevertheless
related to each other by a unitary transformation.  Thus the present study
puts Friar's original considerations in the modern framework of $\chi$EFT
and extends them to the TPE sector.  In Sec.~\ref{sec:hint} and App.~\ref{app:l2piN}
we list those terms in the chiral Lagrangians (and corresponding interaction
Hamiltonians) that are relevant to our purpose here.  In Sec.~\ref{sec:road}
we provide an overview of the derivation of the OPE and TPE potentials and
charge operators.  Specifically, we explore the connection between the amplitude
calculated in $\chi$EFT and the strong and electromagnetic potentials, which
are derived from it and are used in quantum-mechanical formulations, based on
the Lippmann-Schwinger or Schr\"odinger equations.  Power counting
allows us to establish a criterion to make this connection precise.
Elsewhere~\cite{Pastore08,Pastore09}, we have referred to the latter
as ``accounting for recoil-corrected reducible contributions.''
The present formulation is especially apt to shed light on the
inter-dependence between charge operators and potentials (at the
OPE and TPE level), and the associated ambiguity arising from
off-the-energy-shell extrapolations prescribed for the latter.

In Sec.~\ref{sec:road} we also provide an explicit expression for
the unitary transformation, and show, in particular, that different
(non-static) versions of the TPE potential are unitarily equivalent.
As mentioned above, in the OPE sector this result has been
known for a \hbox{while~\cite{Friar77,Adam93}.}

Section~\ref{sec:cnt3} contains a summary of the derivation
of the two-nucleon charge operators up to order $e\, Q^{\, 0}$ included, or
next-to-next-to-next-to-leading order (N3LO in short), where
$e$ is the proton electric charge and $Q$ denotes generically
the low-momentum scale.  No loops enter at this order.  The
different forms of the OPE charge operator at N3LO exhibit
the same unitary equivalence as the non-static corrections
to the OPE potential at order $Q^2$ (N2LO)---this too is well
known~\cite{Friar77,Adam93}, albeit in a different context.

In Sec.~\ref{sec:cnt4} we discuss the static one-loop corrections
to the charge operator at N4LO ($e\, Q$).   In particular,
we list those corresponding to two different off-the-energy-shell
prescriptions for the OPE and TPE potentials, and show that they
are unitarily equivalent.  A fairly detailed account of their
derivation is provided in App.~\ref{app:rho_star}.  The loop
integrals entering the individual terms at N4LO are ultraviolet
divergent; however, their sum is finite, in particular it vanishes
in the limit in which the momentum ${\bf q}$ carried by the
electromagnetic field is zero.  This was to be expected, since i)
symmetry arguments prevent the presence of counterterms
at this order (and therefore the possibility of re-absorbing
ensuing divergencies into them), and ii) charge conservation demands that
at ${\bf q}=0$ the charge operator merely counts the number of
charged particles ({\it i.e.}, protons) in the system---a requirement
already fulfilled at LO.  Summary and conclusions are presented in
Sec.~\ref{sec:con}, while, for future convenience, we give the configuration
space representation of the N4LO charge operators in App.~\ref{app:rspace}.

\section{Relevant interaction Hamiltonians}
\label{sec:hint}

Here we only list the interaction Hamiltonians relevant for the derivation of
the nuclear electromagnetic charge operator up to order $e\, Q$ (N4LO), see
App.~\ref{app:l2piN} for notation and a summary of the corresponding 
Lagrangians~\cite{Bernard95,Fettes98}:
\begin{eqnarray}
 \!H_{\pi N} \!\!&=&\!\!\!\int\!\!\!{\rm d}{\bf x}
 N^\dagger\Big[ \frac{g_A\,\tau_a}{F_\pi}
 {\bm \sigma} \cdot {\bm \nabla} \pi_a \!+\!\frac{{\bm \tau}}{F_\pi^2}
\cdot
({\bm \pi}\times \partial^{\, 0} {\bm \pi}) \Big]\!N  ,
\label{eq:hpiN} \\
H_{\gamma N}\!\! &=&\!\!  e\int\!\!{\rm d}{\bf x}\, N^\dagger \Big[ e_N \, A^0  
 - \frac{2\, \mu_N - e_N }{8\, m_N^2} 
\big[ ({\bm \nabla }^2 A^0)\nonumber \\
& +& {\bm \sigma} \times  ({\bm \nabla} A^0) \cdot \overrightarrow{\bm \nabla}
- \overleftarrow{\bm \nabla} \cdot {\bm \sigma} \times  ({\bm \nabla} A^0) \big]  \Big] N \ ,
\label{eq:hgn} \\
H_{\gamma \pi} \!\!&=&\!\! e\int\!\!{\rm d}{\bf x}\,
 A^0\, \left( {\bm \pi} \times  \partial_0 {\bm \pi} \right)_z \ ,
\label{eq:hgpi}\\
 H_{ \gamma \pi N}\!\! &=&\!\! e \int\!\!{\rm d}{\bf x}\, N^\dagger\,
{\bm \sigma}\cdot\left( {\bm \nabla} A^0\right)\,
\Big[ \frac{ g_A}{2\, m_N F_\pi}
 \left( {\bm \tau}\cdot {\bm \pi}+\pi_z \right)
 \nonumber \\
&+&\frac{1}{F_\pi}\,(2\, d_{20}+ 2\, d_{21}-d_{22}) \,({\bm \tau}\times \partial^{\, 0}{\bm \pi})_z \,
 \Big]\,  N \ .
\label{eq:hgpin}
\end{eqnarray}
The resulting vertices behave, relative to the low-momentum
scale $Q$, in the following way: $H_{\pi N}\sim Q $; first term in $H_{\gamma N} \sim e\, Q^{\,0}$,
remaining ones $\sim e\, Q^2$; $H_{\gamma \pi} \sim e\, Q$; first term in
$H_{\gamma \pi N} \sim e\, Q$, second one $ \sim e\, Q^2$.
There is also a contact interaction,
\begin{eqnarray}
H_{\rm CT}&=&\frac{1}{2}\int{\rm d}{\bf x}\, \Big [ 
 C_S \left(N^\dagger N\right) \left(N^\dagger N\right)\nonumber\\
&+&C_T \left(N^\dagger {\bm \sigma} N\right)\cdot \left(N^\dagger{\bm \sigma} N\right) \Big ]\ ,
\end{eqnarray}
which enters the derivation of the N4LO charge operator.  The accompanying
vertex scales as $Q^{\, 0}$.
\section{From amplitudes to potentials}
\label{sec:road}

We begin by considering the conventional perturbative expansion
for the two-nucleon ($NN$) scattering amplitude
\begin{equation}
 \langle f \mid T\mid i \rangle= 
 \langle f \mid H_1 \sum_{n=1}^\infty \left( 
 \frac{1}{E_i -H_0 +i\, \eta } H_1 \right)^{n-1} \mid i \rangle \ ,
\label{eq:pt}
\end{equation}
where $\mid i \rangle$ and $\mid f \rangle$ represent the initial and final
two-nucleon states of energy $E_i=E_f$, $H_0$ is the Hamiltonian
describing free pions and nucleons, and $H_1$ is the Hamiltonian
describing interactions among these particles (Sec.~\ref{sec:hint}).  The evaluation of this amplitude
is carried out in practice by inserting complete sets of $H_0$ eigenstates
between successive terms of $H_1$.   Power counting is then used
to organize the expansion in powers of $Q/\Lambda_\chi \ll 1 $, where
$\Lambda_\chi \simeq 1$ GeV is the typical hadronic mass scale.

In the perturbative series, Eq.~(\ref{eq:pt}), a generic (reducible or
irreducible) contribution is characterized by a certain number, say
$N$, of vertices, each scaling as $Q^{\alpha_i}\times Q^{-\beta_i/2}$
($i$=$1,\dots,N$), where $\alpha_i$ is the power counting implied by the
relevant interaction Hamiltonian and $\beta_i$ is the number of
pions in and/or out of the vertex, a corresponding $N$--1 number of energy
denominators, and possibly $L$ loops~\cite{Pastore08}.  Out of these
$N$--1 energy denominators, $N_K$ of them will involve only nucleon kinetic
energies, which scale as $Q^2$, and the remaining $N-N_K-1$ will involve,
in addition, pion energies, which are of order $Q$.  Loops, on the other hand,
contribute a factor $Q^3$ each, since they imply integrations
over intermediate three momenta.  Hence the power counting
associated with such a contribution is
\begin{equation}
\left(\prod_{i=1}^N  Q^{\alpha_i-\beta_i/2}
\right)\times \left[ Q^{-(N-N_K-1)}\, Q^{-2N_K} \right ]
\times Q^{3L} \ .
\label{eq:count}
\end{equation}
Clearly, each of the $N-N_K-1$ energy denominators can be further expanded as
\begin{eqnarray}
\frac{1}{E_i-E_I-\omega_\pi}= -\frac{1}{\omega_\pi}
\bigg[ 1 &+& \frac{E_i-E_I}{\omega_\pi}\nonumber\\
&+&
\frac{(E_i-E_I)^2}{\omega^2_\pi} + \dots\bigg] \ ,
\label{eq:deno}
\end{eqnarray}
where $E_I$ denotes the kinetic energy of the intermediate two-nucleon state, and
$\omega_\pi$ the pion energy (or energies, as the case may be)---the ratio $(E_i-E_I)/\omega_\pi$ is of order $Q$.

\begin{figure}[bth]
\includegraphics[width=3.3in]{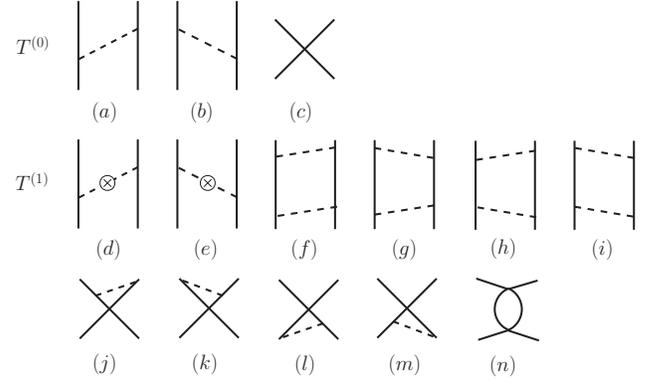}\\
\caption{Time ordered diagrams illustrating the contributions to the
$T^{(0)}$, panels (a)--(c), and $T^{(1)}$, panels (d)--(n),
$N N$ scattering amplitudes.  Nucleons and pions are denoted by
solid and dashed lines, respectively.  Pion lines (pion lines with
crossed circle) indicate that only the leading $Q^{\,-1}$
(next-to-leading $Q^{\,0}$) term in the expansion
of energy denominators, Eq.~(\ref{eq:deno}), are retained in the
corresponding amplitudes. See text for explanation.}
\label{fig:f1}
\end{figure}

The $Q$-scaling of the interaction vertices
and the considerations above show that $T$ admits the following expansion:
\begin{equation}
 T=T^{(0)} + T^{(1)} + T^{(2)} + \dots \ ,
\label{eq:tmae}
\end{equation}
where $T^{(n)} \sim Q^n$.  For example, the time-ordered diagrams contributing to
$T^{(0)}$ and $T^{(1)}$ are illustrated in Fig.~\ref{fig:f1}, where the
pion line (pion line with a crossed circle) indicates that only the leading
$-1/\omega_\pi$ (next-to-leading $-(E_i-E_I)/\omega^2_\pi$) term is retained in the expansion of the associated energy denominator, Eq.~(\ref{eq:deno}).  Except for App.~\ref{app:rho_star}, this notation will not be used any further below, but it is understood that energy denominators involving pions are expanded as in Eq.~(\ref{eq:deno}).

Our objective is to derive a two-nucleon potential $v$ which, when iterated in the
Lippmann-Schwinger (LS) equation,
\begin{equation}
v+v\, G_0\, v+v\, G_0 \, v\, G_0 \, v +\dots \ ,
\label{eq:lse}
\end{equation}
leads to the $T$-matrix in Eq.~(\ref{eq:tmae}), order by order in the power counting.
In practice, this requirement can only be satisfied up to a given order $n^*$, and
the resulting potential, when inserted into the LS (or Schr\"odinger) equation, will
generate contributions of order $n > n^*$, which do not match $T^{(n)}$.
In Eq.~(\ref{eq:lse}), $G_0$ denotes the free two-nucleon propagator, $G_0=1/(E_i-E_I+i\, \eta)$,
and we assume that
\begin{equation}
v=v^{(0)}+v^{(1)}+v^{(2)}+\dots\ ,
\end{equation}
where the yet to be determined $v^{(n)}$ is of order $Q^n$.  We also note that, generally, a term
like $v^{(m)}\, G_0 \, v^{(n)}$ is of order $Q^{m+n+1}$, since $G_0$ is of order $Q^{-2}$
and the implicit loop integration brings in a factor $Q^3$.  Having established the above power counting,
we obtain
\begin{eqnarray}
v^{(0)} &=& T^{(0)} \ , \label{eq:v0}\\
v^{(1)} &=& T^{(1)}-\left[ v^{(0)}\, G_0\, v^{(0)}\right] \ , \\
v^{(2)} &=& T^{(2)}-\left[ v^{(0)}\, G_0\, v^{(0)}\, G_0\, v^{(0)}\right] \nonumber\\
&&\qquad-\left[ v^{(1)}\, G_0 \, v^{(0)}
+v^{(0)}\, G_0\, v^{(1)}\right] \ , \label{eq:v2}\\
v^{(3)} &=& T^{(3)}-\left[v^{(0)}\, G_0\, v^{(0)}\, G_0\, v^{(0)}\, G_0\, v^{(0)}\right]
\nonumber \\
&&\qquad-
\left[v^{(1)} \, G_0\, v^{(0)} \, G_0\, v^{(0)}+ {\rm permutations}\right] \nonumber \\
&&\qquad -\left[ v^{(2)}\, G_0\, v^{(0)} +  v^{(0)}\, G_0\, v^{(2)}\right] \nonumber\\
&&\qquad-
\left[v^{(1)} \, G_0\, v^{(1)}\right] \ , \label{eq:v3}
\end{eqnarray}
where $v^{(n)}$ is the ``recoil-corrected'' two-nucleon potential explicitly
constructed in Refs.~\cite{Pastore08,Pastore09} up to order $n=2$,
or N2LO.  The LO term, $v^{(0)}$,
consists of (static) one-pion-exchange (OPE) and contact interactions, while
the NLO term, $v^{(1)}$, vanishes,
since the contributions of diagrams (d) and (e)  in $T^{(1)}$, illustrated
in Fig.~\ref{fig:f1}, add up to zero, while the remaining diagrams represent iterations
of $v^{(0)}$, whose contributions are exactly canceled by
$\left[ v^{(0)}\, G_0\, v^{(0)}\right]$---complete or partial cancellations of this type
persist at higher ($n\ge 2$) orders.
The N2LO term, which follows from Eq.~(\ref{eq:v2}), contains two-pion-exchange (TPE) and contact
(involving two gradients of the nucleon fields) interactions.  It is derived in Ref.~\cite{Pastore09}.
However, there is a recoil correction of order $n=2$ to the OPE potential, which was ignored in that
paper.  In momentum space, it is given by
\begin{equation}
v^{(2)}_\pi(\nu=0) = v^{(0)}_\pi({\bf k}) \,\,\frac{(E_1^{\, \prime}-E_1)^2+(E_2^{\, \prime}-E_2)^2}{2\, \omega_k^2}  \ ,
\label{eq:vpi2}
\end{equation}
where $v^{(0)}_\pi$ is the LO OPE potential,
\begin{equation}
\label{eq:vpi0}
v^{(0)}_\pi({\bf k})=-\frac{g_A^2}{F_\pi^2}\, {\bm \tau}_1\cdot {\bm \tau}_2\,
 \frac{{\bm \sigma}_1 \cdot {\bf k}\,\,{\bm \sigma}_2\cdot {\bf k}}{\omega_k^2} \ ,
\end{equation}
${\bf k}={\bf p}_1-{\bf p}^\prime_1={\bf p}_2^\prime-{\bf p}_2$ is the momentum transfer,
and ${\bf p}_j$ and $E_j$ (${\bf p}^\prime_j$ and $E_j^{\, \prime}$) are the initial (final) 
momentum and energy of \hbox{nucleon $j$}.
Obviously, on the energy shell $E_i=E_f$ implicit in Eq.~(\ref{eq:v2}), the
above expression is equivalent to one in which, for example,
\begin{equation}
v^{(2)}_\pi(\nu=1) =-v^{(0)}_\pi({\bf k}) \,\,\frac{(E_1^{\, \prime}-E_1)\,
(E_2^{\, \prime}-E_2)}{\omega_k^2}  \ .
\label{eq:vpi2n}
\end{equation}
In fact, there is an infinite class of $v_\pi^{(2)}(\nu)$ corrections---labeled by the
parameter $\nu$~\cite{Friar77,Friar80,Adam93}---which, while equivalent on the energy-shell,
are different off the energy-shell, and therefore lead to different potentials $v^{(3)}(\nu)$
in Eq.~(\ref{eq:v3}).  Indeed, for the choices of $v_\pi^{(2)}(\nu)$ in Eqs.~(\ref{eq:vpi2})
and~(\ref{eq:vpi2n}) the corresponding corrections to the TPE term (from direct and
crossed box diagrams), $v^{(3)}_{{2\pi}}(\nu)$, read
\begin{eqnarray}
\label{eq:v3box}
v_{{2 \pi}}^{(3)}(\nu=0)\!\!&=&\!\!
-\frac{g_A^4}{2\,F_\pi^4} (3+2\,{\bm \tau}_1\cdot{\bm \tau}_2)
\int _{{\bf q}_1}({\bm \sigma}_1\cdot {\bf q}_2)({\bm \sigma}_1\cdot{\bf q}_1)
\nonumber\\
&\times&\!\!
({\bm \sigma}_2\cdot{\bf q}_1)({\bm \sigma}_2\cdot{\bf q}_2)
\bigg( \frac{E_1-\widetilde{E}_1+E_2^\prime-\widetilde{E}_2^\prime}
{\omega_1^4\,\omega_2^2}\nonumber\\
&+&\!\! \frac{E_1^\prime-\widetilde{E}_1+E_2-\widetilde{E}_2^\prime}
{\omega_1^2\,\omega_2^4} \bigg) \ ,
\end{eqnarray}
and
\begin{eqnarray}
\!\!\!v_{{2 \pi}}^{(3)}(\nu=1)\!\!&=&\!\! v_{{2 \pi}}^{(3)}(\nu=0) 
+\frac{1}{2}\int_{{\bf q}_1} v_\pi^{(0)}({\bf q}_2)\, v_\pi^{(0)}({\bf q}_1)\nonumber \\
&\times& \!\!\!\left( E_1+E_2-\widetilde{E}_1 -\widetilde{E}_2\right)\!
\left( \frac{1}{\omega_1^2}+\frac{1}{\omega_2^2} \right)  ,
\label{eq:v3nu}
\end{eqnarray}
where, as indicated in panel (a) of Fig.~\ref{fig:fig4} (App.~\ref{app:rho_star}),
${\bf q}_1$ and ${\bf q}_2$ ($\omega_1$ and $\omega_2$) are
the momenta (energies) of the two exchanged
pions (with ${\bf q}_2={\bf k}-{\bf q}_1$), $\widetilde{E}_j$
and $\widetilde{E}_j^{\, \prime}$
are the intermediate nucleon energies,
and
\begin{equation}
\int_{{\bf s}}\equiv\int\frac{{\rm d}{\bf s}}{(2\pi)^3} \ .
\end{equation}

Friar~\cite{Friar77,Friar80}, and later Adam {\it et al.}~\cite{Adam93},
have argued that the different off-the-energy-shell extrapolations
$v^{(2)}_\pi(\nu)$ are unitarily equivalent.  (See also
Ref.~\cite{Coon88} for the implication of the unitary
equivalence on the TPE three-nucleon potential.) 
We show below that, as a matter of fact,
this unitary equivalence remains valid for $v^{(3)}_{2\pi}(\nu)$ (as well as for the
electromagnetic charge operators at one loop), thus extending the results
of the authors of Refs.~\cite{Friar80,Adam93} to the TPE sector.
Up to order $n=3$ ({\it i.e.}, $Q^3$), the two-nucleon Hamiltonian
in the center-of-mass (CM) frame can be written in momentum space as
\begin{eqnarray}
H(\nu;{\bf p}^\prime,{\bf p})\!\!&=&\!\!K^{(-1)}({\bf p}^\prime,
{\bf p})+v_\pi^{(0)}({\bf p}^\prime\!-{\bf p})
+v_{2\pi}^{(2)}({\bf p}^\prime-{\bf p})\nonumber\\
&+&\!\!v_\pi^{(2)}(\nu;{\bf p}^\prime,{\bf p})+v_{2\pi}^{(3)}
(\nu;{\bf p}^\prime,{\bf p}) \ ,
\label{eq:hnnu}
\end{eqnarray}
limiting our considerations to OPE and (box) TPE potentials only.  There are, of course,
$v^{(3)}$ terms originating from higher order chiral
Lagrangians~\cite{Bernard95,Fettes98}, but these have no relevance
for the discussion to follow, and are therefore ignored below.  In Eq.~(\ref{eq:hnnu}), $K^{(-1)}$ denotes
the kinetic energy term of order $n=-1$,
\begin{equation}
 K^{(-1)}({\bf p}^\prime,{\bf p})=(2\pi)^3\, \delta({\bf p}^\prime-{\bf p})\, p^2/m_N \ ,
\end{equation}
$m_N$ is the nucleon mass, and
$v^{(2)}_\pi(\nu)$ has been given in Refs.~\cite{Friar80,Adam93} as
\begin{equation}
v_\pi^{(2)}(\nu;{\bf p}^\prime,{\bf p})=(1-2\, \nu)\,
\frac{v_\pi^{(0)}({\bf p}^\prime-{\bf p})}{({\bf p}^\prime-
{\bf p})^2+m_\pi^2}\, \frac{(p^{\prime\, 2}-p^{\, 2})^2}{4\, m_N^2}\ ,
\label{eq:vpifr}
\end{equation}
which the $\nu=0,1$ expressions listed above reduce to (in the CM frame).
These Hamiltonians are related to each other via
\begin{equation}
H(\nu)={\rm e}^{-iU(\nu)}\, H(\nu=0)\,{\rm e}^{+iU(\nu)} \ ,
\end{equation}
where up to NLO the operator $i\, U(\nu)$ is
\begin{equation}
i\,U(\nu;{\bf p}^\prime,{\bf p})\simeq i\, U^{(0)}(\nu;{\bf p}^\prime,{\bf p})
+i\, U^{(1)}(\nu;{\bf p}^\prime,{\bf p}) \ , 
\end{equation}
with
\begin{eqnarray}
\label{eq:uu0}
i\,U^{(0)}(\nu;{\bf p}^\prime,{\bf p})\!\!&=&\!\!-\nu\,
\frac{v_\pi^{(0)}({\bf p}^\prime-{\bf p})}{({\bf p}^\prime-
{\bf p})^2+m_\pi^2}\, \frac{p^{\prime\, 2}-p^{\, 2}}{2\, m_N} \ ,\\
\label{eq:uu1}
i\,U^{(1)}(\nu;{\bf p}^\prime ,{\bf p})\!\!&=&\!\!
- \frac{\nu}{2} \int_{\bf s} \frac{v_\pi^{(0)}({\bf p}^\prime-{\bf s}) v_\pi^{(0)}({\bf s}-{\bf p})}
{({\bf p}^\prime-{\bf s})^2+m_\pi^2}\ .
\end{eqnarray}
The unitary equivalence up to order $n=3$ implies
\begin{eqnarray}
H(\nu)=H(\nu=0)&+&\left[ K^{(-1)}+v^{(0)}_\pi\, , \, i\, U^{(0)}(\nu) \right] \nonumber\\
&+&
\left[ K^{(-1)} , \, i\, U^{(1)}(\nu) \right]\ ,
\end{eqnarray}
since each commutator brings in an additional factor $Q^{\,3}$ due to
the implicit momentum integrations.  A direct evaluation
with $\nu=1$ shows that $H(\nu=1)$ ensues, including $v_\pi^{(2)}(\nu=1)$ and
$v^{(3)}_{2\pi}(\nu=1)$ as given in Eqs.~(\ref{eq:vpi2n}) and~(\ref{eq:v3nu})---note
that in the CM frame ${\bf q}_1={\bf p}-{\bf s}$, ${\bf q}_2={\bf s}-{\bf p}^\prime$,
$\widetilde{E}_1+\widetilde{E}_2=s^2/m_N$, and ${\bf s}$ is the loop momentum.
Both these $\nu$-dependent corrections are relevant for the derivation of the nuclear
charge operator up to N4LO, to which we now turn our attention.

The electromagnetic interactions are treated in first order
in the perturbative expansion of Eq.~(\ref{eq:pt}), and
the transition operator can be expanded as
\begin{equation}
T_\gamma=T_\gamma^{(-3)}+T_\gamma^{(-2)}+T_\gamma^{(-1)} +\dots \ ,
\end{equation}
where $T_\gamma^{(n)}$ is of order $e\, Q^n$ ($e$ is the electric charge).
The nuclear charge, $\rho$, and current, ${\bf j}$, operators follow
from $v_\gamma= A^0\, \rho-{\bf A}\cdot {\bf j}$, where
$A^\mu=(A^0,{\bf A})$ is the electromagnetic vector
field, and it is assumed that $v_\gamma$ has a similar expansion as $T_\gamma$.
The requirement that, in the context of the LS equation,
$v_\gamma$ matches $T_\gamma$ order by order in the power counting
implies the following relations:
\begin{eqnarray}
v_\gamma^{(-3)}= T_\gamma^{(-3)}\!\!\!\!\!\!\!
&& \, \label{eq:vg3m} \\
v_\gamma^{(-2)}= T_\gamma^{(-2)}\!\!\!\!\!\!\!
&&-\left[ v_\gamma^{(-3)}\, G_0\, v^{(0)}+
v^{(0)}\, G_0\, v_\gamma^{(-3)} \right] \ , \\
v_\gamma^{(-1)}=T_\gamma^{(-1)}\!\!\!\!\!\!\!
&&-\left[ v_\gamma^{(-3)}\, G_0\, v^{(0)}\, G_0\, v^{(0)}
+{\rm permutations} \right] \nonumber \\
\!\!\!\!\!\!\!&&-
\left[v_\gamma^{(-2)}\, G_0\, v^{(0)}+v^{(0)}\, G_0\, v_\gamma^{(-2)}\right]  \ ,
\\
\label{eq:vg0}
v_\gamma^{(0)}= T_\gamma^{(0)}\!\!\!\!\!\!\!
&&-\,\Big[v_\gamma^{(-3)}\, G_0\, v^{(0)}\, G_0\, v^{(0)}\, G_0\, v^{(0)}\nonumber\\
\!\!\!\!\!\!\!&&
\,+\,{\rm permutations}\Big] \nonumber \\
\!\!\!\!\!\!\!&&-\left[v_\gamma^{(-2)} \, G_0\, v^{(0)} \, G_0\, v^{(0)}+ {\rm permutations}\right]\nonumber\\
\!\!\!\!\!\!\!&&-\left[ v_\gamma^{(-1)}\, G_0\, v^{(0)} +
v^{(0)}\, G_0\, v_\gamma^{(-1)}\right] \nonumber \\
\!\!\!\!\!\!\!&&-\left[ v_\gamma^{(-3)}\, G_0\, v^{(2)}+ v^{(2)}\, G_0\, v_\gamma^{(-3)}\right] \ ,
\end{eqnarray}
\begin{eqnarray}
v_\gamma^{(1)}= T_\gamma^{(1)}\!\!\!\!\!\!\!&&-
\,\Big[v_\gamma^{(-3)}\, G_0\, v^{(0)}\, G_0\, v^{(0)}\, G_0\, v^{(0)}\, G_0\,
v^{(0)}\nonumber\\
\!\!\!\!\!\!\!&&+ \,{\rm permutations}\Big]\nonumber\\
\!\!\!\!\!\!\!&&-\,\Big[v_\gamma^{(-2)}\, G_0\, v^{(0)}\, G_0\, v^{(0)}\, G_0\, v^{(0)}
\nonumber\\
\!\!\!\!\!\!\!&&+ \,{\rm permutations}\Big] \nonumber \\
\!\!\!\!\!\!\!&&-\left[v_\gamma^{(-1)}\, G_0\, v^{(0)}\, G_0\, v^{(0)}+{\rm permutations}\right]\nonumber\\
\!\!\!\!\!\!\!&&-
\left[v_\gamma^{(0)}\, G_0\, v^{(0)}+v^{(0)}\, G_0\, v_\gamma^{(0)}\right] \nonumber \\
\!\!\!\!\!\!\!&&-
 \left[v_\gamma^{(-3)}\, G_0\, v^{(2)}\, G_0\, v^{(0)}+{\rm permutations}\right]\nonumber\\
\!\!\!\!\!\!\!&&-
\left[v_\gamma^{(-3)}\, G_0\, v^{(3)}+ v^{(3)} \, G_0\, v_\gamma^{(-3)}\right] \ ,
\label{eq:vg1}
\end{eqnarray}
where $v_\gamma^{(n)}=A^0\, \rho^{(n)}-{\bf A}\cdot {\bf j}^{(n)}$,
$v^{(n)}$ are the $NN$ potentials constructed in Eqs.~(\ref{eq:v0})--(\ref{eq:v3})
(with the $\nu$ dependence of $v^{(2)}$ and $v^{(3)}$ suppressed for the time being),
and use has been made of the fact that $v^{(1)}$ vanishes.  In the propagator $G_0$,
the initial energy $E_i$ includes the photon energy $\omega_\gamma$ (itself of order
$Q^2$), {\it i.e.}~$E_i=E_1+E_2+\omega_\gamma=E_1^{\, \prime}+E_2^{\, \prime}$,
and the intermediate energy $E_I$ may include, in addition to the kinetic energies
of the intermediate nucleons, also the photon energy, depending on the specific
time ordering being considered.

The current operators ${\bf j}^{(n)}$ up to order $n=1$, {\it i.e.} $e\, Q$, have
been derived in Ref.~\cite{Pastore09}.  In that case, the derivation is fairly straightforward
as ${\bf j}^{(-3)}$ vanishes: the lowest order ($n=-2$) contributing to ${\bf j}$
consists of the single-nucleon convection and spin-magnetization currents.  The situation
for the charge operator is considerably more complicated, however, since $n=-3$
is the lowest order contributing to it---in momentum space, it is given by
\begin{equation}
\rho^{(-3)}({\bf q})=e\, e_{N,1}\, (2\pi)^3\delta({\bf p}_1+{\bf q}-{\bf p}_1^\prime) + 1 \rightleftharpoons 2 \ ,
\label{eq:r-3}
\end{equation}
where $e_{N,i}=( 1+\tau_{i,z} )/2$ is the proton projection operator, ${\bf q}$ is
the momentum carried by the external field, and the counting
$e\, Q^{-3}$ follows from the product of a factor $e\, Q^0$ associated with the
$\gamma NN$ vertex, and a factor $Q^{-3}$ due to the momentum-conserving
$\delta$-function implicit in a disconnected term of this type, see panel (a) in Fig.~\ref{fig:fig1}
below.  Therefore, the operators $\rho^{(0)}$ and
$\rho^{(1)}$, obtained from Eqs.~(\ref{eq:vg0}) and~(\ref{eq:vg1}), depend on the
off-the-energy-shell extensions adopted for $v^{(2)}$ and $v^{(3)}$.  In particular,
it appears that all of these extensions lead to a $\rho^{(1)}$ operator which i) is free
of divergencies, as required by the absence of counterterms at this order
($e\, Q$), and ii) satisfies $\rho^{(1)}({\bf q}=0)=0$.  This last condition
follows from charge conservation,
\begin{equation}
\rho({\bf q}=0)=\int{\rm d}{\bf x}\, \rho({\bf x}) =\rho^{(-3)}({\bf q}=0) \ ,
\end{equation}
implying $\rho^{(n\ge -2)}({\bf q}=0)=0$.  In Sec.~\ref{sec:cnt4} and App.~\ref{app:rho_star},
we show explicitly that the off-the-energy-shell prescriptions adopted for
$v^{(2)}(\nu)$ and $v^{(3)}(\nu)$ corresponding to $\nu=0,1$
do ensure that $\rho^{(1)}(\nu;{\bf q})$ obeys requirements i)  and ii).
Indeed, we also show that the unitary equivalence extends to the
OPE $\rho^{(0)}(\nu;{\bf q})$---a fact already known~\cite{Friar77,Friar80}---and
TPE $\rho^{(1)}(\nu;{\bf q})$ charge operators.

\section{Charge operators up to N3LO}
\label{sec:cnt3}

\begin{figure}[bth]
\includegraphics[width=3.3in]{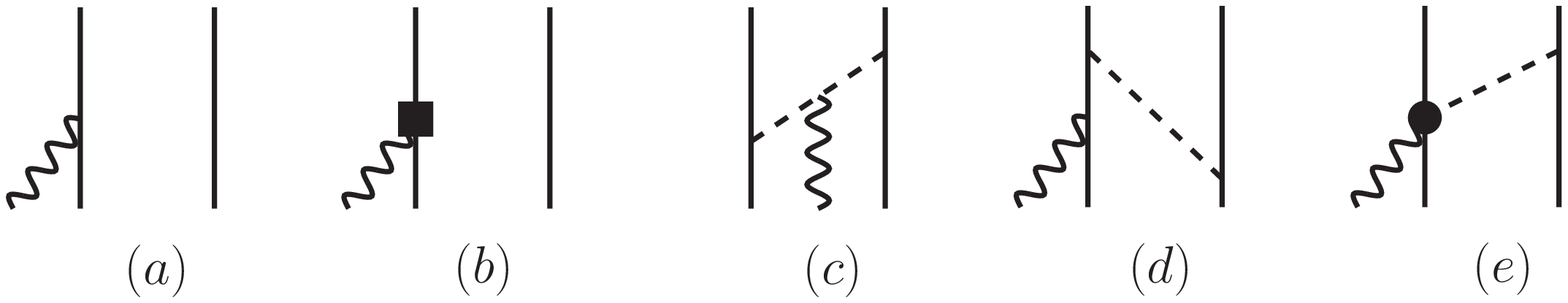}\\
\caption{Diagrams illustrating one- and two-body charge operators entering at LO ($e\, Q^{-3}$),
panel (a), N2LO ($e\, Q^{-1}$), panels (b), (c) and (d), and N3LO ($e\, Q^{0}$), panels (e).
There are no NLO contributions.  Nucleons, pions, and photons are denoted by solid, dashed,
and wavy lines, respectively.  The square in panel (b) represents the $(Q/m_N)^2$, or $(v/c)^2$, relativistic
correction to the LO one-body charge operator, whereas the solid circle in panel (e) is associated
with a $\gamma \pi N$ charge coupling of order $e\, Q$ (see text).  Only one among the possible
time orderings is shown in panels (c), (d), and (e).}
\label{fig:fig1}
\end{figure}

The LO contribution to the two-nucleon charge operator in panel (a)
of Fig.~\ref{fig:fig1}, resulting from the first
term of the $\gamma N$ interaction Hamiltonian in Eq.~(\ref{eq:hgn}),
has already been given in Eq.~(\ref{eq:r-3}).
There are no NLO ($e\, Q^{-2}$) contributions, whereas at N2LO there is i)
a relativistic correction of order $(Q/m_N)^2$ to the LO charge operator, panel (b), given by
\begin{eqnarray}
\rho^{(-1)}&=& -\frac{e}{8 \, m_N^2} \, \left( 2\, \mu_{N,1}-e_{N,1}\right)\nonumber\\
&\times&\big(q^2
+2\, i\, {\bf q}\cdot {\bm \sigma}_1\times {\bf K }_1\big)
 + 1 \rightleftharpoons 2  \ ,
\label{eq:r1rc}
\end{eqnarray}
ii) a pion-in-flight term, panel (c), which, however, turns out to vanish when
the contributions of the six time-ordered diagrams, evaluated in the static limit,
are summed up, iii) a one-pion-exchange (OPE) contribution, panel (d), which vanishes due to an
exact cancellation between static irreducible and recoil corrected reducible amplitudes~\cite{Pastore08}.
In Eq.~(\ref{eq:r1rc}) and what follows, ${\bf q}$ denotes the
momentum carried by the external field, and ${\bf k}_i$ and ${\bf K}_i$ are defined as
\begin{equation}
{\bf k}_i = {\bf p}_i^\prime -{\bf p}_i \ , \qquad {\bf K}_i =({\bf p}_i^\prime +{\bf p}_i)/2 \ ,
\label{eq:defs}
\end{equation}
where ${\bf p}_i$ and ${\bf p}_i^\prime$ are the initial and final momenta of nucleon $i$.
Hereafter, momentum conserving $\delta$-functions (${\bf q}={\bf k}_i$) in $\rho^{(-3)}$ and $\rho^{(-1)}$
(and ${\bf k}_1+{\bf k}_2={\bf q}$ in the following expressions of two-body charge operators)
will be dropped for brevity.
We note that the power counting is different for the electromagnetic
current operator, for which the LO term is of order $e\, Q^{-2}$ (in the
two-nucleon system), {\it i.e.}~it is suppressed by an extra power of $Q$
relative to $\rho^{(-3)}$, and where there are NLO ($e\, Q^{-1}$) corrections
involving seagull and in-flight contributions associated with OPE,
which have no counterpart in the present case.

The N3LO contribution illustrated in panel (e) of Fig.~\ref{fig:fig1}
is associated with the $\gamma \pi N$ coupling of order $e\, Q$ originating
from the first term in Eq.~(\ref{eq:hgpin}) ;
it gives rise to the vertex
\begin{equation}
i\, \frac{e\, g_A}{2\, m_N F_\pi}\, \frac{{\bm \sigma}\cdot {\bf q}}{\sqrt{2\omega_k}}
\, (\tau_a + \delta_{az})
\end{equation}
for absorption (or emission) of a pion of momentum ${\bf k}$, energy $\omega_k$, and
isospin component $a$, where $(2\omega_k)^{-1/2}$ is
the normalization factor entering the normal modes expansion of the pion field.
The two-body charge operator follows easily by evaluating (in the static
limit) the contributions of the two time-ordered diagrams:
\begin{equation}
\rho^{(0)}_{\rm e}=\frac{e}{2\, m_N} \frac{g_A^2}{F_\pi^2} \left( {\bm \tau}_1 \cdot {\bm \tau_2}
+ \tau_{2z}\right)\, \frac{{\bm \sigma}_1 \cdot {\bf q} \,\, {\bm \sigma}_2 \cdot {\bf k}_2}{\omega^2_{k_2}}
+ 1 \rightleftharpoons 2 \ .
\label{eq:pich}
\end{equation}
In the present $\chi$EFT context, $\rho^{(0)}_{\rm e}$ was derived first by
Phillips in 2003~\cite{Phillips03}.  However, it is worthwhile to point out that the
presence of an operator of the form given
in Eq.~(\ref{eq:pich}) has been known for some time---see the 1989 review paper
by Riska~\cite{Riska89} and references therein.  It was obtained by considering
the low-energy limit of the relativistic Born diagrams associated
with the virtual-pion photoproduction amplitude.  Subsequently,
calculations based on realistic wave functions for the $A=2$--4 nuclei
showed that this operator plays an important role
in yielding predictions for the $A$ structure function and tensor polarization of the deuteron~\cite{Schiavilla91},
and charge form factors of the trinucleons and $\alpha$-particle~\cite{Schiavilla90}, that
are in excellent agreement with the experimental data at low and moderate values of
the momentum transfer ($q \lesssim 1$ GeV/c).   These calculations also showed that
the contributions due to $\rho^{(0)}_{\rm e}$ are typically an order of
magnitude larger than those generated by the Darwin-Foldy and spin-orbit
relativistic corrections---{\it i.e.}, the operator $\rho^{(-1)}$ above---or
by vector-meson exchanges.

There are also N3LO contributions originating from non-static
contributions in diagrams of type (c) and (d), resulting
from expanding the energy denominators involving pions as in
Eq.~(\ref{eq:deno}).  We obtain:
\begin{eqnarray}
\rho_{\rm c}^{(0)}&=&i \frac{e}{m_N}\, \frac{g_A^2}{F_\pi^2}
\left({\bm \tau}_1\times{\bm \tau}_2\right)_z
\frac{{\bm \sigma}_1\cdot{\bf k}_1\,{\bm \sigma}_2
\cdot{\bf k}_2}{\omega_{k_1}^2\,\omega_{k_2}^2} \nonumber\\
&\times&\left({\bf k}_1\cdot{\bf K}_1-{\bf k}_2\cdot{\bf K}_2\right) \ .
\label{eq:rho_gpipi}
\end{eqnarray}
The contributions from diagrams of type (d) depend on the off-the-energy-shell prescription adopted for
$v^{(2)}_\pi(\nu)$~\cite{Friar77,Adam93,Friar80}.  For $\nu=0$ and $\nu=1$, we find by
direct evaluation of the relevant diagrams:
\begin{eqnarray}
\label{eq:rhon0}
\rho^{(0)}_{\rm d}(\nu=0)\!\!\!&=&\!\!\!-\frac{e}{4\, m_N}\, \frac{g_A^2}{F_\pi^2}
\frac{{\bm \sigma}_1\cdot{\bf k}_2\,{\bm \sigma}_2
\cdot{\bf k}_2}{\omega_{k_2}^4}\nonumber\\
&\times&\!\!\!\Big[
\left( {\bm \tau}_1\cdot{\bm \tau}_2+\tau_{2,z}\right)
 {\bf q}\cdot {\bf k}_2+2\, i\,
\left( {\bm \tau}_1\times{\bm \tau}_2\right)_z \nonumber\\
&\times&\!\!\!{\bf k}_2 \cdot ({\bf K}_1+{\bf K}_2)\Big]+1 \rightleftharpoons 2 , \\
\label{eq:rhon1}
\rho^{(0)}_{\rm d}(\nu=1)\!\!\!&=&\!\!\!-i\, \frac{e}{m_N}\, \frac{g_A^2}{F_\pi^2}\,
\left( {\bm \tau}_1\times{\bm \tau}_2\right)_z
\frac{{\bm \sigma}_1\cdot{\bf k}_2\,{\bm \sigma}_2\cdot{\bf k}_2}{\omega_{k_2}^4}
\nonumber\\
&\times&\!\!\! {\bf k}_2 \cdot {\bf K}_2+1 \rightleftharpoons 2 \ ,
\end{eqnarray}
and it is easily seen that they are related to each other by the unitary transformation $U^{(0)}(\nu)$, that is
\begin{eqnarray}
\rho^{(0)}_{\rm d}(\nu)&=&
\rho^{(0)}_{\rm d}(\nu=0)+\left[ \rho^{(-3)}\, , \, i\, U^{(0)}(\nu) \right]\nonumber\\
&=& \rho^{(0)}_{\rm d}(\nu=0) +
i\, e\,\Big[ e_{N,1}\, U^{(0)}(\nu;{\bf p}^\prime 
-{\bf q}/2,{\bf p}) \nonumber \\
&-&U^{(0)}(\nu;{\bf p}^\prime,{\bf p}+{\bf q}/2)\,e_{N,1}\Big] + 1\rightleftharpoons 2\ ,
\label{eq:uni0}
\end{eqnarray}
and ${\bf p}$ and ${\bf p}^\prime$ are the initial and final relative momenta.
We observe that $\rho_{\rm c}^{(0)}+\rho_{\rm d}^{(0)}(\nu)=0$ in the limit
${\bf q}=0$, as required by charge conservation.  We also point out that
the isovector term proportional to $({\bm \tau}_1\times{\bm \tau}_2)_z$
in Eq.~(\ref{eq:rhon0}) vanishes in the Breit frame,
where ${\bf p}_1+{\bf p}_2=-{\bf q}/2$ and
${\bf p}_1^\prime+{\bf p}_2^\prime={\bf q}/2$.

\section{Charge operator at N4LO}
\label{sec:cnt4}
\begin{figure}[bth]
\includegraphics[width=3.3in]{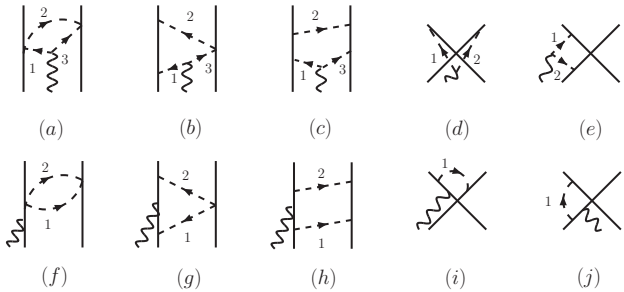}\\
\caption{Diagrams illustrating one--loop charge operators entering at N$^4$LO ($e\, Q$),
notation is as in Fig.~\ref{fig:fig1}. Only one among the possible time orderings is shown for each contribution.}
\label{fig:fig2}
\end{figure}

First, we note that there are non-static N4LO contributions
from diagrams of type (c)--(e) in Fig.~\ref{fig:fig1}---though,
those relative to panel (e) cancel out, when the two time orderings
are taken into account.  There is also a non-static contribution of
order $(Q/m_N)^4$ to the LO $\rho^{(-3)}$ charge operator, which we ignore in
the present section.  Here we only deal with static N4LO corrections
from one-loop diagrams of the type represented in Fig.~\ref{fig:fig2},
since those induced by the second term in the interaction $H_{\gamma\pi N}$
(proportional to the time derivative of the pion field) vanish.
Thus up to N4LO included, there are no unknown low-energy
constants entering the electromagnetic charge operator.

The pion-in-flight contributions illustrated in panels (a) and (b)
of Fig.~\ref{fig:fig2} involve irreducible
diagrams only, and they are obtained by direct evaluation,
in the static limit, of the corresponding amplitudes.
We find that the ``football'' contribution---panel (a)---vanishes,
while the ``triangle'' pion-in-flight operator---panel (b)---reads
\begin{eqnarray}
\label{eq:rho_b}
 \rho^{(1)}_{\rm b}&=&e\,\frac{2\,g_A^2}{F_\pi^4}\, \tau_{2,z}
\int \frac{{\bf q}_1\cdot{\bf q}_2}{\omega_1^2\,\omega_2^2} +1\rightleftharpoons2 \ ,
\end{eqnarray}
where the ${\bf q}_i$ and $\omega_i=(q_i^2+m_\pi^2)^{1/2}$ denote the momenta
(with the flow as indicated in the figure) and energies of the exchanged pions,
and the integration is on any one of the ${\bf q}_i$'s, the remaining
${\bf q}_j$'s with $j\not= i$ being fixed by momentum-conserving $\delta$-functions---as
noted in the previous section, an overall $(2\pi)^3\delta({\bf k}_1+{\bf k}_2-{\bf q})$
has been dropped.

Diagrams illustrated in panels (c)--(j) of Fig.~\ref{fig:fig2}
have both reducible and irreducible pieces.  As discussed in Sec.~\ref{sec:road},
the evaluation of the amplitudes is carried
out retaining recoil corrections to the reducible diagrams
(up to N4LO accuracy, in this particular instance),
along with the static (N4LO) irreducible contributions.
We find that recoil corrected reducible contributions partially
cancel static irreducible terms at the same order.  This is discussed
in considerable detail in App.~\ref{app:rho_star}.
Consequently, at N4LO the charge operator associated with diagrams of type (c)
shown in Fig.~\ref{fig:fig2} reads
\begin{eqnarray}
\label{eq:rho_c}
\rho^{(1)}_{\rm c}&=& \!\!- e\,\frac{2\,g_A^4}{F_\pi^4}
\int \frac{1}{\omega_1^2\,\omega_2^2\,\omega_3^2}
\Big[2\left( \tau_{1,z}+\tau_{2,z}\right)\nonumber\\
&\times&\!\!\left({\bf q}_2\cdot{\bf q}_1\,{\bf q}_2\cdot{\bf q}_3-
{\bm \sigma}_1\cdot{\bf q}_2\times{\bf q}_1\,{\bm \sigma}_2\cdot{\bf q}_3\times{\bf q}_2\right)\nonumber\\
&-&\!\!\left({\bm \tau}_1\times{\bm \tau}_2\right)_z
({\bf q}_1\cdot{\bf q}_2\,{\bm \sigma}_2\cdot{\bf q}_3\times{\bf q}_2
\nonumber\\
&+&\!\!
{\bf q}_2\cdot{\bf q}_3\,{\bm \sigma}_1\cdot{\bf q}_2\times{\bf q}_1)\Big] \ ,
\end{eqnarray}
while that arising from contributions of type (d) diagrams
vanishes, since the integrand is an odd function of the loop momentum
${\bf q}_1$.
For type (e) diagrams, we find:
\begin{eqnarray}
\label{eq:rho_e}
\rho_{\rm e}^{(1)}&=&\!\!\,e \frac{2 g_A^2}{F_{\pi}^2}\left(\tau_{1,z}+\tau_{2,z}\right)\,\int
\, \frac{1}{\omega_1^2\,\omega_2^2}\Big[\,C_S \, {\bf q}_1\cdot{\bf q}_2\nonumber\\
&+&\!\!C_T\,( {\bm \sigma}_1\cdot{\bf q}_1\,{\bm \sigma}_2\cdot{\bf q}_2
+{\bm \sigma}_1\cdot{\bf q}_2 \,{\bm \sigma}_2\cdot{\bf q}_1
\nonumber\\&-&\!\!
{\bf q}_1\cdot{\bf q}_2\,{\bm \sigma}_1\cdot{\bm \sigma}_2  ) \Big] \ .
\end{eqnarray}
The $\rho^{(1)}_{\rm f}$ operator vanishes due to an exact cancellation
between the static irreducible and recoil-corrected
reducible amplitudes associated with the diagrams illustrated in
panel (f).  For type (g)--(j) diagrams, we find:
\begin{eqnarray}
\label{eq:rho_f}
\rho^{\rm (1)}_{\rm g}\!\!&=&\!\!- e \,\frac{2 \, g_A^2}{F_\pi^4}\,\tau_{2,z}\!
\int \frac{{\bf q}_1\cdot{\bf q}_2}{\omega^2_1\,\omega_2^2}+1\rightleftharpoons2  , \\
 \label{eq:rho_g}
\rho_{\rm h}^{(1)}(\nu=0)\!\!&=&\!\!-e \,\frac{ 2\,g_A^4}{F_{\pi}^4}\,\int
\frac{\omega_1^2+\omega_2^2}
{\omega_1^4\,\omega_2^4}
\Big[\tau_{2,z}\,
({\bf q}_1\cdot{\bf q}_2)^2 \nonumber \\
&+&\!\!\tau_{1,z}\,
{\bm \sigma}_2\cdot{\bf q}_2\times{\bf q}_1 
\,{\bm \sigma}_1\cdot{\bf q}_2\times{\bf q}_1
  \Big] \nonumber \\
&-&\!\! e \,\frac{ g_A^4}{F_{\pi}^4}\,({\bm \tau}_1\times{\bm \tau}_2)_z\,\int
\frac{\omega_1^2-\omega_2^2}
{\omega_1^4\,\omega_2^4}\, {\bf q}_1\cdot{\bf q}_2\nonumber\\
&\times&\!\!{\bm \sigma}_1\cdot{\bf q}_2\times{\bf q}_1+1\rightleftharpoons2 \ ,\\
\label{eq:rho_gn1}
 \rho_{\rm h}^{(1)}(\nu=1)\!\!&=&\!\!\rho_{\rm h}^{(1)}(\nu=0)
+\Bigg[i\,e \frac{g_A^4}{2\,F_\pi^4}\,({\bm \tau}_1\times{\bm \tau}_2)_z
\nonumber\\
&\times&\!\!\int \bigg[\frac{\omega_1^2+\omega_2^2}{\omega_1^4\, \omega_2^4}
\big[{\bm \sigma}_1\cdot{\bf q}_2\times{\bf q}_1\,
{\bm \sigma}_2\cdot{\bf q}_2\times{\bf q}_1 
\nonumber\\
&-&\!\!({\bf q}_1\cdot{\bf q}_2)^2
\big] - i \, \frac{\omega_1^2-\omega_2^2}{\omega_1^4\, \omega_2^4}\,
{\bf q}_1\cdot{\bf q}_2\nonumber\\
&\times&\!\!\left({\bm \sigma}_1 + {\bm \sigma}_2\right)
\cdot{\bf q}_2\times{\bf q}_1\bigg]
+1\rightleftharpoons2 \,\Bigg]\ ,\\
\rho^{(1)}_{\rm i}(\nu=0)\!\!&=&\rho^{(1)}_{\rm i}(\nu=1)= 0 \ ,\\
\label{eq:rho_i}
\rho_{\rm j}^{(1)}\!\!&=&\!\!e\,\frac{2\, g_A^2}{3\,F_{\pi}^2}\,\tau_{1,z} 
\left(3\, C_S-C_T \,{\bm \sigma}_1\cdot{\bm \sigma}_2\right)\nonumber\\
&\times&\!\!
\int \frac{q_1^2}
{\omega_1^4}\,+1\rightleftharpoons2 \ ,
\end{eqnarray}
and a fairly detailed overview of their derivation is
in App.~\ref{app:rho_star}.

A few comments are now in order. Firstly, the loop integrals entering
the expressions above are ultra-violet divergent.  However,
the total charge operator at N4LO is finite, since the divergencies
associated with contributions (b) and (g), (c) and (h), and (e) and (j)
cancel out.  This is in line with the fact that there are no
counterterms at this order.  In particular, we observe that the constraint imposed
by charge conservation is satisfied, since $\rho^{(1)}_{\rm b}+\rho^{(1)}_{\rm g}=0$
and $\rho^{(1)}_{\rm e}+\rho^{(1)}_{\rm j}=0$ in the limit ${\bf q}=0$
(or ${\bf k}_1=-{\bf k}_2$), while
the contribution associated with diagram (c) in Fig.~\ref{fig:fig2} can
be written as (since ${\bf q}_1=-{\bf q}_3$ at ${\bf q}=0$)
 \begin{eqnarray}
\label{eq:rho_cr}
\rho^{(1)}_{\rm c}({\bf q}=0)&=& \!\!e\,\frac{2\,g_A^4}{F_\pi^4}
\int_{{\bf q}_1,{\bf q}_2} \frac{1}{\omega_1^4\,\omega_2^2}
\Big[ 2\, \tau_{2,z}\, ({\bf q}_2\cdot{\bf q}_1)^2
\nonumber\\
&+& 2\, \tau_{1,z}\,
{\bm \sigma}_1\cdot{\bf q}_2\times{\bf q}_1\,\,{\bm \sigma}_2\cdot{\bf q}_2\times{\bf q}_1\nonumber\\
&-&\left({\bm \tau}_1\times{\bm \tau}_2\right)_z
{\bf q}_2\cdot{\bf q}_1\,{\bm \sigma}_1\cdot{\bf q}_2\times{\bf q}_1\Big]
\nonumber\\
&\times& (2\pi)^3\,\delta
({\bf q}_1-{\bf q}_2 -{\bf k}_1) + 1 \rightleftharpoons 2\ .
\end{eqnarray}
It is then seen that, in this limit, the expression above is opposite
in sign to that of diagram (h) in Eq.~(\ref{eq:rho_g})
for $\nu=0$.  For $\nu=1$, in Eq.~(\ref{eq:rho_gn1}) the extra terms
proportional to $({\bm \tau}_1\times{\bm \tau}_2)_z$
vanish by themselves at ${\bf q}=0$.
For completeness, we list the configuration-space representation of these operators
in App.~\ref{app:rspace}.

Secondly, the charge operators $\rho^{(1)}_{\rm h}$ for $\nu=0,1$
are related to each other by the unitary transformation $U$, {\it i.e.}~a relation
similar to Eq.~(\ref{eq:uni0}) holds with $U^{(0)}(\nu)$ being replaced
by $U^{(1)}(\nu)$, defined in Eq.~(\ref{eq:uu1}).  This is easily verified
by expressing $U^{(1)}(\nu)$ as
\begin{eqnarray}
\!\!\!\!\!&&\!\!\!\!\!i\, U^{(1)}(\nu;{\bf p}^\prime-{\bf p})=- \nu\, 
\frac{g_A^4}{4\, F_\pi^4}\,(3/2-{\bm \tau}_1\cdot {\bm \tau}_2)\nonumber \\
\!\!\!\!\!&&\!\!\!\!\! \times\!\! \int_{\bf s}\Bigg[  \frac{\omega_+^2+\omega_-^2}
{\omega_+^4\,\omega_-^4}
\Big[ \left[({\bf p}^\prime-{\bf p})^2-s^2\right]^2\!\!
-\!4 \left[{\bm \sigma}_1\cdot({\bf p}^\prime\!-\!{\bf p})\times
{\bf s}\right] \nonumber\\
\!\!\!\!\!&&\!\!\!\!\!\times
 \left[{\bm \sigma}_2\cdot({\bf p}^\prime\!-\!{\bf p})\times
{\bf s}\right]\Big] -2\, i\, \frac{\omega_-^2-\omega_+^2}
{\omega_+^4\,\omega_-^4} 
\left[({\bf p}^\prime-{\bf p})^2-s^2\right] \nonumber \\
\!\!\!\!\!&&\!\!\!\!\!\times \left( {\bm \sigma}_1+{\bm \sigma}_2\right)
\cdot({\bf p}^\prime-{\bf p})\times {\bf s}
\Bigg] \ ,
\label{eq:ur1}
\end{eqnarray}
where
\begin{equation}
\omega_{\pm} \equiv \sqrt{
({\bf p}^\prime-{\bf p}\pm {\bf s})^2+4\, m_\pi^2} \ .
\label{eq:omg}
\end{equation}
The commutator $\left[ \rho^{(-3)}\, ,\, U^{(1)}(\nu)\right]$ is seen
to be identical to the $({\bm \tau}_1\times{\bm \tau}_2)_z$ term
on the right-hand-side of Eq.~(\ref{eq:rho_gn1}), when the pion
momenta ${\bf q}_{1,2}$ are expressed as ${\bf q}_{1,2}=
{\bf q}/2+{\bf p}-{\bf p}^\prime \pm {\bf s}$
and ${\bf s}$ is the loop momentum.

Thirdly, we compared the operators given above with those
derived by K\"olling {\it et al.}~\cite{Koelling09}
in TOPT with the Okubo method~\cite{Okubo54}, in order to decouple, in the Hilbert space of pions
and nucleons, the states consisting of nucleons only from those
including, in addition, pions.  We find that the
expressions for operators (a), (b), (c), (g), and (h, $\nu=0$) are identical to
those reported in Ref.~\cite{Koelling09}---the terms involving contact interactions in
panels (d), (e), (i, $\nu$), and (j) were not considered by the authors of that paper.
We should note that in $\rho^{(1)}_{\rm h}(\nu=0$)
the additional isovector piece, {\it i.e.}~the
term multiplied by $({\bm \tau}_1\times{\bm \tau}_2)_z$
in Eq.~(\ref{eq:rho_g}), is missing in Ref.~\cite{Koelling09}.
However, evaluation of the loop integral shows that it
vanishes.  Indeed, consider
\begin{eqnarray}
&&\hspace{-0.5cm}\int_{{\bf q}_1} \frac{{\bf q}_1\cdot({\bf k}_2-{\bf q}_1)\,
{\bm \sigma}_1 \cdot({\bf k}_2\times{\bf q}_1)}{(q_1^2+m_\pi^2)\,
[ \left({\bf k}_2-{\bf q}_1\right)^2+m_\pi^2]^2}
 =\int_0^1{\rm d}x\, 2\, x\,(1-2\, x) \nonumber \\
&&\times\!\int_{{\bf q}_1} \frac{{\bf q}_1\cdot{\bf k}_2\,\,
{\bm \sigma}_1 \cdot({\bf k}_2\times{\bf q}_1)}
{\left[q_1^2+m_\pi^2+k_2^2\, x\,(1-x) \right]^3}=0\ ,
\end{eqnarray}
after making use of Feynman's parametrization, and shifting the integration
variables as ${\bf q}_1 -x\, {\bf k}_2 \rightarrow {\bf q}_1$.
Thus, the type (h) charge operator derived in Ref.~\cite{Koelling09}
corresponds to the $\nu=0$ off-the-energy-shell extension.
On the other hand, the framework used by these authors
leads to vanishing non-static corrections to the OPE potential~\cite{Epelbaum05}
(see also the discussion by Phillips~\cite{Phillips07} in connection
to this issue), which would imply the choice $\nu=1/2$ in
Eq.~(\ref{eq:vpifr}).  This suggests that pion retardation effects
may not have been treated consistently in Ref.~\cite{Koelling09}.
We conclude by observing that for clarity's sake we have
kept the (vanishing) isovector terms in $\rho_{\rm h}^{(1)}(\nu)$,
Eqs.~(\ref{eq:rho_g}) and~(\ref{eq:rho_gn1}).

\section{Conclusions}
\label{sec:con}

We have presented a fairly systematic derivation of the
two-nucleon charge operators up to one loop (or N4LO)
in $\chi$EFT, based on TOPT with a careful treatment of
the non-iterative contributions extracted from reducible diagrams.
The specific form of the N3LO and N4LO charge operators depends
on the off-the-energy-shell prescriptions adopted for the non-static
pieces in the OPE and TPE potentials.  This ambiguity is of no
import, however, since these OPE and TPE (non-static) potentials
and accompanying charge operators are related to each other by a unitary
transformation.  Thus, provided a consistent set is adopted, predictions
for physical observables, such as the few-nucleon charge form factors,
will remain unaffected by the non-uniqueness associated with
off-the-energy-shell effects.

However, it is important to stress that in the present work we have only examined
those off-the-energy-shell effects relating to pion retardation~\cite{Friar77,Friar80},
which arise, in TOPT amplitudes, from energy denominators containing pion
(in addition to nucleon kinetic) energies.  There are, of course, additional
non-static corrections originating from the non-relativistic reduction of
interaction vertices (generated by fully relativistic Lagrangians).
Corrections of this type in the OPE sector for both potentials and
charge operators have been studied in Refs.~\cite{Friar77,Adam93,Friar80}.
It would be interesting to extend those considerations to the TPE
sector, and also explore the constraints, in the present  $\chi$EFT
setting, that relativistic covariance and power counting impose on
these non-static terms of the potentials and electromagnetic charge
and current operators.  As a matter of fact, a study along these lines,
but dealing only with the two-nucleon potential, is that of
Ref.~\cite{Girlanda10a}.

Finally, we note that the charge operators up to N4LO included contain
no unknown low-energy constants.
The N4LO corrections are purely isovector,
and will not contribute to isoscalar observables, such as the $A$ structure
function and tensor polarization of the deuteron or charge form factor
of $^4$He.  They will produce, presumably tiny, contributions to the
isovector combination of the trinucleon radii and charge form factors.
A quantitative analysis of all these effects is in progress.

\section*{Acknowledgments}

We would like to thank D.R.\ Phillips for correspondence in reference to his derivation
of the one-pion-exchange charge operator in Eq.~(\ref{eq:pich}).  An interesting
conversation with E.\ Epelbaum, S.\ K\"olling, and H.\ Krebs is also acknowledged
by one of the authors (R.S.).  R.S.~thanks the Physics Department of
the University of Pisa, the INFN Pisa
branch, and especially the Pisa group for the support and warm hospitality
extended to him on several occasions.  His work is supported by the
U.S.~Department of Energy, Office of Nuclear Physics, under contract
DE-AC05-06OR23177.
%
%
%
%
\appendix
\section{Chiral Lagrangians}
\label{app:l2piN}
In the heavy-baryon formalism~\cite{Bernard95,Fettes98}, the chiral
Lagrangians describing the interactions among nucleons, pions, and photons are written as
\begin{eqnarray}
{\cal L}^{(1)}_{\pi N} &=& N^\dagger \left ( i\, v_\mu D^\mu +g_A\, S^\mu u_\mu \right) N \ ,
\label{eq:l1} \\
{\cal L}^{(2)}_{\pi N} &=&\frac{1}{2\, m_N} N^\dagger \Big[
v_\mu \,v_\nu\, D^\mu D^\nu -D^\mu  D_\mu\nonumber\\
&-&i \, g_A\, S^\mu  v^\nu
\left[  D_\mu\, ,\, u_\nu \right]_+ \nonumber \\
& - &e\, \mu_N \, \epsilon^{\mu \nu \rho \sigma}\, F_{\mu \nu} v_\rho \, S_\sigma +\dots \Big]N \ ,
\label{eq:l2} \\
{\cal L}^{(3)}_{\pi N} &=&N^\dagger \Bigg[i\,e\,d_{20}\, S^\mu v^\nu
F_{\mu\nu} \left[ \tau_z \, ,\, u_\rho \right] v^\rho\nonumber\\
&+&i\,e\,d_{21}\, S^\mu F_{\mu \nu} \left[ \tau_z \, ,\, u^\nu \right]
+ e\,d_{22}\, S^\mu\, \left[D^\nu\, , \,  F^-_{\mu\nu} \right]\nonumber\\
&+&e \left(2\, d_7+d_6 \,\tau_z
 -  \frac{2\, \mu_N - e_N}{8\, m_N^2} \right) \left[ D^\mu\, ,\, F_{\mu \nu} \right] v^\nu        \nonumber \\
&-& e\, \frac{2\, \mu_N - e_N}{4\, m_N^2}
( i \,\epsilon^{\mu\nu\alpha\beta}\, v_\alpha\, S_\beta\, F_{\mu \sigma}\, v^\sigma D_\nu
\nonumber\\
 &+& {\mathrm{h.c.}})+\dots \Bigg]N 
\label{eq:l3} \\
{\cal L}^{(2)}_{\pi \pi}&=& \frac{F_\pi^2}{16}\, {\rm tr} \left[ D_\mu U\,  D^\mu U^\dagger
+m_\pi^2\left( U+U^\dagger \right) \right] \ ,
\label{eq:lp}
\end{eqnarray}
where the fields $U$ and $u_\mu$, and the covariant derivatives $D^\mu N$ and $D^\mu U$, are
given by
\begin{eqnarray}
\!\!\!\!\!\!\!\!\!\!\!\!
U+U^\dagger &=&2 -\frac{4}{F_\pi^2} {\bm \pi} \cdot {\bm \pi} + \dots \ , \\
u_\mu &=& -\frac{2}{F_\pi} {\bm \tau} \cdot \partial_\mu {\bm \pi}-\frac{2\, e}{F_\pi}\, A_\mu\,
 ( {\bm \tau} \times {\bm \pi} )_z + \dots \ ,\\
D_\mu N&=& \bigg[ \partial_\mu +i \, e \, e_N A_\mu + \frac{i}{F_\pi^2} {\bm \tau}
\cdot ( {\bm \pi}\times \partial_\mu {\bm \pi})\nonumber\\
&-&\frac{i\, e}{F_\pi^2} A_\mu\,
\left[ {\bm \pi} \times ( {\bm \tau} \times {\bm \pi})\right]_z + \dots \bigg] N \ , \\
D_\mu U &=& \frac{2\, i}{F_\pi} \, {\bm \tau }\cdot \partial_\mu {\bm \pi}-\frac{4}{F_\pi^2}
{\bm \pi} \cdot \partial_\mu {\bm \pi}\nonumber\\
 &+&\frac{2\, i\, e}{F_\pi} A_\mu ({\bm \tau} \times {\bm \pi})_z + \dots \ ,\\
F^-_{\mu \nu} &=& \frac{2\, e}{F_\pi}\, F_{\mu \nu} \, \left( {\bm \tau}\times {\bm \pi}\right)_z + \dots \ ,
\end{eqnarray}
$g_A $, $F_\pi$ ($F_\pi=186$ MeV), and $e$ are, respectively, the nucleon axial coupling constant,
pion decay amplitude, and proton electric charge, $d_6$, $d_7$, $d_{20}$, $d_{21}$, and 
$d_{22}$ are (unknown) low-energy constants (LEC's),  $\left[ \dots\, ,\, \dots\right]_+$
denotes the anticommutator, and
$v^\mu$ and $S^\mu =(i/2) \, \gamma_5\, \sigma^{\mu \nu}\, v_\nu$ are
the nucleon's four-velocity and spin operator, which
in its rest frame reduce to $v^\mu = (1, {\bm 0})$ and $S^\mu= (0, {\bm \sigma}/2)$.
We have also defined
\begin{equation}
e_N = (1+\tau_z)/2 \ , \,\,\,
\kappa_N =  (\kappa_S+ \kappa_V \, \tau_z)/2 \ , \,\,\, \mu_N = e_N+\kappa_N  \ ,
\label{eq:ekm}
\end{equation}
where $\kappa_S$ and $\kappa_V$ are the isoscalar and isovector combinations
of the anomalous magnetic moments of the proton and neutron, which are
related to the LEC's $c_6$ and $c_7$ used in Ref.~\cite{Fettes98}
as $\kappa_S = c_6 + 2\, c_7$ and $\kappa_V = c_6$.
Note that only electromagnetic interaction terms have been
included in ${\cal L}^{(3)}_{\pi N}$, and that the terms proportional
to the LEC's $d_6$ and $d_7$ represent corrections arising from
the nucleon substructure ({\it i.e.}, an electromagnetic
form factor).  They are not relevant to our discussion here, and
will be ignored hereafter, together with the effects due to the pion cloud of the nucleon. 
In the expressions above, the electromagnetic vector and tensor fields are denoted by
$A^\mu$ and $F^{\mu\nu}$, the isospin doublet of (nonrelativistic) nucleon fields by $N$, and
the isospin triplet of pion fields by ${\bm \pi}$.  In terms of these, the Lagrangians are
expressed as
\begin{eqnarray}
\!\!\!\!\!\!\!\!\!{\cal L}^{(1)}_{\pi N} \!\!&=&\!\! N^\dagger \Big [ i \, \partial^{\,0} - \frac{g_A}{F_\pi}\, \tau_a\,
 {\bm \sigma} \cdot {\bm \nabla} \pi_a -\frac{1}{F_\pi^2} {\bm \tau} \cdot
({\bm \pi}\times \partial^{\, 0} {\bm \pi}) \nonumber\\
&-& \!\! e\, e_N \, A^0 +
\frac{e}{F_\pi^2}\, A^0 \,\left[ {\bm \pi} \times ({\bm \tau} \times {\bm \pi} )\right]_z 
 +\dots \Big]N
 \label{eq:lpn1} \ , \\
{\cal L}^{(2)}_{\pi N}\!\! &=&\!\! \frac{1}{2\, m_N}N^\dagger \Big [ \nabla^2
-\frac{e\, g_A}{F_\pi} \left( {\bm \tau}\cdot {\bm \pi}+\pi_z \right)
 {\bm \sigma}\nonumber
\\
&\cdot&\!\! \left({\bm \nabla}A^0 \right) +
\dots \Big] N  \ , 
\label{eq:lpin2} \\
{\cal L}^{(3)}_{\pi N}\!\! &=&\!\!
N^\dagger \Big[ -\frac{e\,(2\,d_{20}+2\, d_{21}-d_{22})}{F_\pi}
{\bm \sigma}\cdot\left( {\bm \nabla} A^0 \right)
({\bm \tau}\times \partial^{\, 0}{\bm \pi})_z\nonumber\\
&+&\!\!\frac{e\,(2 \, \mu_N - e_N)}{8\, m_N^2}\big[ \left({\bm \nabla}^2 A^0\right)+
{\bm \sigma} \cdot  ({\bm \nabla} A^0) \times \overrightarrow{\bm \nabla}
\nonumber\\& +& 
{\bm \sigma} \cdot \overleftarrow{\bm \nabla} \times ({\bm \nabla} A^0) \big]
+\dots \Big] N\ ,
\label{eq:lpin3} \\
{\cal L}^{(2)}_{\pi \pi}\!\! &=&\!\!
 \frac{1}{2}\, \partial_\mu {\bm \pi} \cdot \partial^\mu {\bm \pi}
-\frac{m_\pi^2}{2}\, {\bm \pi }\cdot {\bm \pi}
\nonumber\\
&-&\!\!e\, A^\mu\, ({\bm \pi}\times \partial_\mu {\bm \pi})_z
+\dots \ ,
\label{eq:lpp2}
\end{eqnarray}
where the term proportional to $e \, g_A/F_\pi$ in the second line of Eq.~(\ref{eq:lpin2}) is obtained~\cite{Phillips03}
\hbox{i) by expanding} the anticommutator in Eq.~(\ref{eq:l2}) as
\begin{eqnarray}
-i \frac{g_A}{2\, m_N} N^\dagger\,  S^\mu  v^\nu 
\left[  D_\mu\, ,\, u_\nu \right]_+ N &=& i \frac{g_A}{ m_N \, F_\pi}\, N^\dagger \, S^\mu  v^\nu\nonumber\\
&&\!\!\!\!\!\!\!\!\!\!\!\!\!\!\!
\!\!\!\!\!\!\!\!\!\!\!\!\!\!\!
\!\!\!\!\!\!\!\!\!\!\!\!\!\!\!
\!\!\!\!\!\!\!\!\!\!\!\!\!\!\!
\!\!\!\!\!\!\!\!\!\!\!\!\!\!\!\times
\Big[  D_\mu\, ,\,  {\bm \tau} \cdot \partial_\nu {\bm \pi}-e\, A_\nu\,
 ( {\bm \tau} \times {\bm \pi} )_z\Big]_+ N \ ,
\end{eqnarray}
ii) by removing the derivative $\partial_\nu$ acting on the pion field via partial integration,
and iii) by using the (lowest order) equation of motion for the nucleon field, {\it i.e.}
\begin{equation}
i\, v^\nu \, \partial_\nu  N= -e\, e_N\,  v^\nu  A_\nu N + \dots \ ,
\end{equation}
to re-express the terms, which result from ii) and involve $i\, v^\nu \, \partial_\nu N$ and
its adjoint.  In Eqs.~(\ref{eq:lpn1})--(\ref{eq:lpp2}), we have retained only linear terms in
the vector potential, and only contributions relevant
for the derivation of the two-body charge operator up to order $e\, Q$.  Application
of the standard rules of canonical quantization leads to the interaction Hamiltonians
listed in Sec.~\ref{sec:hint}.

\section{Derivation of the N4LO Charge Operator}
\label{app:rho_star}

In this appendix, we derive the static N4LO corrections
at one loop to the electromagnetic charge operator
which follow from Eq.~(\ref{eq:vg1}).  The derivation of the operators
associated with the irreducible contributions illustrated
by panels (a) and (b) in Fig.~\ref{fig:fig2} is straightforward.
However, the analysis of the reducible diagrams of type (c)--(j)
in the same figure is more delicate, since the corresponding
amplitudes contain (static) contributions originating from
two different sources: one arising from the sub-class of
irreducible time orderings for each of the diagrams (c)--(j),
and one consisting of the ``left-over'' in the reducible time
orderings, after the energy-dependent terms representing
iterations in the Lippmann-Schwinger (LS) equation have
been properly identified and removed, {\it i.e.}~cancelled
by terms on the right-hand-side of Eq.~(\ref{eq:vg1}).
The latter will be referred to below as ``recoil-corrected''
reducible contributions.
\begin{figure}[bth]
\includegraphics[width=3.3in]{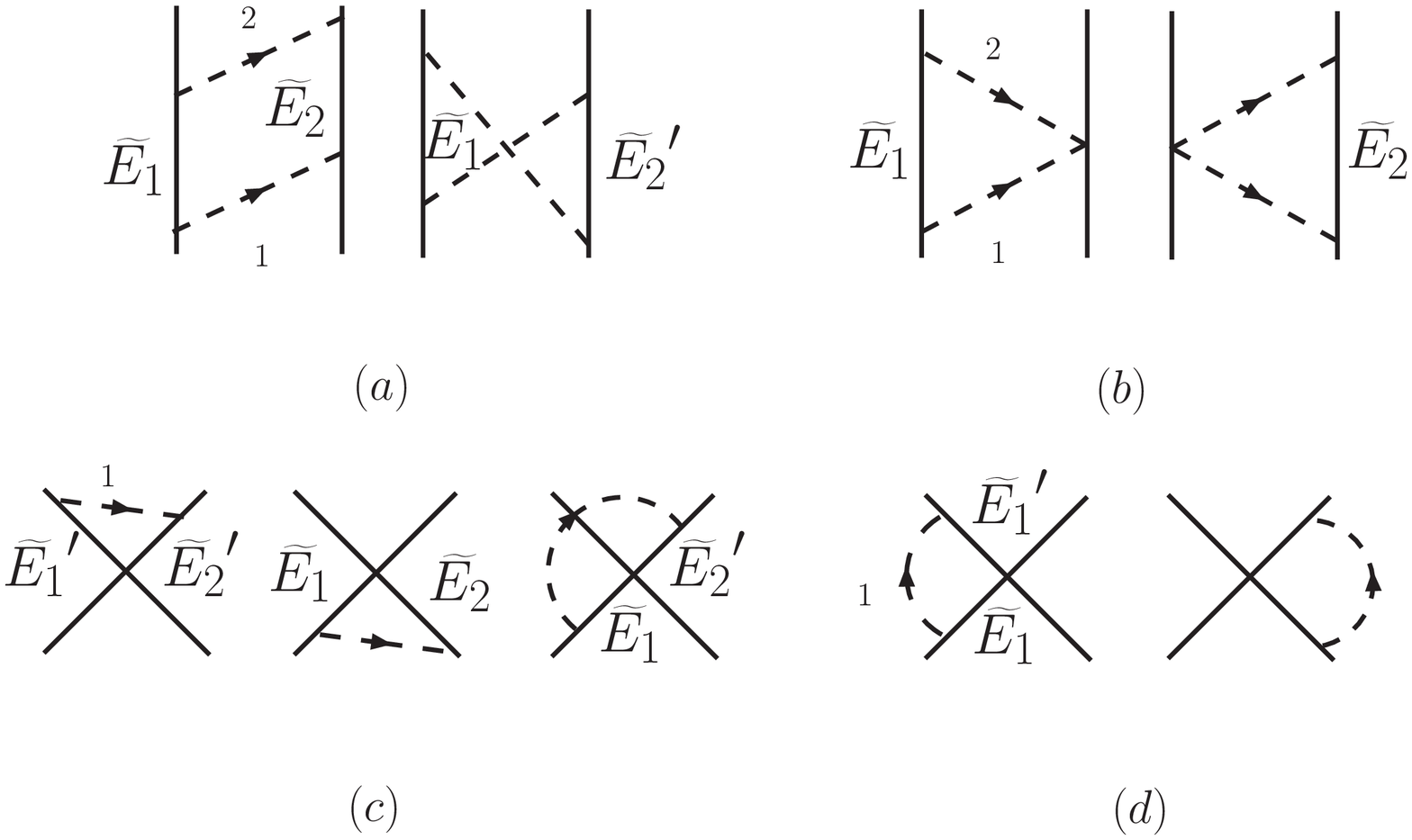}\\
\caption{Diagrams illustrating the
$T^{(3)}$ $N N$ amplitudes. The kinetic energies of intermediate
nucleons are as given.  Only one among the possible
time orderings is shown.  Notation is as in Fig.~\protect\ref{fig:fig1}.}
\label{fig:fig4}
\end{figure}
As mentioned in Sec.~\ref{sec:road}, in order to isolate
these recoil-corrected pieces from those embedded into
the iterated solution of the LS equation, it is necessary
to identify the formal expressions of the N3LO contributions
to the $NN$ potential.  The latter are obtained by retaining
terms beyond the leading one in the expansion of the energy
denominators including pion energies, which enter both
the reducible and irreducible amplitudes.  Note that the N3LO
contributions from higher order chiral Lagrangians are of no
interest here.  Next section
is devoted to the derivation of the N3LO $NN$ potential, while the
last two sections deal with the derivation of the N3LO one-pion-exchange
(OPE) and N4LO two-pion-exchange (TPE) charge operators.

\subsection{$NN$ potential at N3LO}
\label{app:rc_NNpot}

One-loop contributions to the $N N$ potential
considered in this appendix are shown in Fig.~\ref{fig:fig4}.
We discuss in depth the results obtained for the
box diagrams shown in panel (a) of this figure.
The remaining corrections can be easily derived
following the steps outlined here, and for them
we only provide a listing of their expressions.

\begin{figure}[bth]
\includegraphics[width=3.3in]{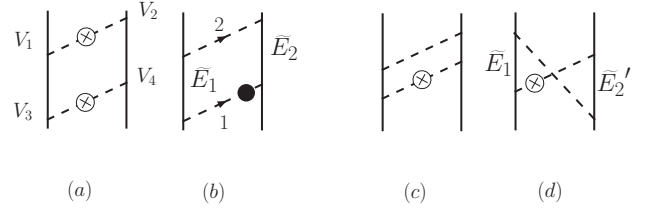}\\
\caption{Diagrams illustrating the recoil-corrected reducible,
panels (a) and (b), and irreducible, panels (c) and (d), amplitudes
contributing to the $N N$ potential at N3LO.
Pion lines with crossed (full) circle indicate that
only the next-to-leading $Q^{\,0}$ (next-to-next-to-leading $Q^{\,1}$)
term in the expansion of energy denominators, Eq.~(\ref{eq:deno}),
are retained in the corresponding amplitudes. See text for explanation.
In panels (b), (c), and (d), the crossed or full circles
can be either on pion one (as shown in the figure) or on pion two.
Only one among the possible time orderings is shown.
Notation is as in Fig.~\protect\ref{fig:fig1}.}
\label{fig:fig5}
\end{figure}

The classes of diagrams contributing to the N3LO box amplitude
are illustrated in Fig.~\ref{fig:fig5}.  The type (a) reducible, and type (c) and
(d) irreducible diagrams, are evaluated by keeping next-to-leading (or $Q^{\, 0}$)
terms in the expansions of the energy denominators which include
pion energies---see Eq.~(\ref{eq:deno}).  The amplitude corresponding
to the reducible type (b) diagrams is obtained by
retaining terms of order $Q$ in these energy denominators, namely
one order higher than for diagrams of type (a), (c), and (d).

We write the N3LO amplitude associated with the box
diagrams as the sum of reducible
and irreducible contributions, {\it i.e.}
\begin{equation}
\label{eq:v2pi}
T_{2\pi}^{(3)}(\nu)=
T_{2\pi,\,\rm red}^{(3)}(\nu)+
T_{2\pi,\,\rm irr}^{(3)} \ .
\end{equation}
The reducible amplitude consists of LS terms plus a term
contributing to the definition of the $N N$ potential at N3LO.
The latter is affected by the choice of the off-the-energy-shell
prescription adopted for the N2LO OPE potential $v^{(2)}_\pi(\nu)$.
As an example, we discuss the results obtained
with $\nu=0$ and $\nu=1$. In particular,
for $\nu=0$, we find that the box reducible amplitude
is given by
\begin{eqnarray}
\label{eq:v2pi_red0}
T_{2\pi,\,\rm red}^{(3)}(\nu=0)&=&
2\,\frac{V_1\,V_2}{\omega_2}
\frac{1}{E_i-\widetilde{E}_1-\widetilde{E}_2}\nonumber\\
&\times&
 \frac{V_3\,V_4}{\omega_1^3} 
\left[(E_1-\widetilde{E}_1)^2+(E_2-\widetilde{E}_2)^2\right]\nonumber\\
&+& \frac{V_1\,V_2}{\omega_2^3}
\Big[(\widetilde{E}_1-E^{\, \prime}_1)^2+(\widetilde{E}_2-E^{\, \prime}_2)^2\Big]\nonumber\\
&\times&
\frac{1}{E_i-\widetilde{E}_1-\widetilde{E}_2}
\,2\,\frac{V_2\,V_3}{\omega_1}\nonumber\\
&+&\,V_1\,V_2\,V_3\,V_4\,
\frac{E_i-\widetilde{E}_1-\widetilde{E}_2}
{\omega_1^2\,\omega_2^2} \ ,
\end{eqnarray}
where $E_i=E_1+E_2=E_1^\prime+E_2^\prime$ is
the initial energy of the system, $\widetilde{E}_j$ and ${\bf q}_i$
($\omega_i$) denote, respectively, the energies  of the intermediate nucleons
and momenta (energies) of the exchanged pions as
indicated in panels (a) and (b) of Fig.~\ref{fig:fig5}, and an integral over an
unconstrained pion momentum is understood.
In the equation above, and through the remainder of this appendix,
we denote with $V_i$ the vertices entering the diagrams.
These vertices are implied by the interaction Hamiltonians listed in
Sec.~\ref{sec:hint}.  For example, the $V_1$ vertex shown in panel
(a) of Fig.~\ref{fig:fig5} is associated with the $H_{\pi N}$ Hamiltonian,
and reads
\begin{equation}
V_1=-i\,\frac{g_A}{F_\pi}
\frac{{\bm \sigma}_1\cdot{\bf q}_2}{\sqrt{2\,\omega_2}}\tau_{1,b} \ ,
\end{equation}
where $b$ specifies the isospin component of the pion.

The last term in Eq.~(\ref{eq:v2pi_red0}) is the
N3LO recoil-corrected reducible contribution to the $NN$ potential
mentioned earlier, corresponding to the prescription $\nu=0$.
After resolving the spin-isospin structure implied by the
vertices, one can easily recognize that the first two terms in $T^{(3)}_{2\pi,{\rm red}}$
represent iterations of the LS equation with the static, $v_\pi^{(0)}$,
and N2LO, $v_\pi^{(2)}(\nu=0)$, OPE potentials defined in Eq.~(\ref{eq:vpi0})
and~(\ref{eq:vpi2}), respectively, namely
\begin{eqnarray}
T_{2\pi,\,\rm red}^{(3)}(\nu=0)\!\!&=&\!\! v_{\pi}^{(0)}\,G_0\,v_{\pi}^{(2)}(\nu=0)
+ v_{\pi}^{(2)}(\nu=0)\,G_0\,v_{\pi}^{(0)} \nonumber\\
&&\!\!+\,V_1\,V_2\,V_3\,V_4\,
\frac{E_i-\widetilde{E}_1-\widetilde{E}_2}
{\omega_1^2\,\omega_2^2} \ ,
\end{eqnarray}
where, for brevity, the dependence upon
nucleon energies and pion momenta is not
explicitly indicated.  It can be inferred from
Fig.~\ref{fig:fig5}.  If the prescription
$\nu=1$ is considered for $v^{(2)}_\pi(\nu)$,
the box reducible amplitude at N3LO reads instead
\begin{eqnarray}
\label{eq:v2pi_red1}
T_{2\pi,\,\rm red}^{(3)}(\nu=1)\!\!&=&\!\! v_{\pi}^{(0)}\,G_0\,v_{\pi}^{(2)}(\nu=1)
+ v_{\pi}^{(2)}(\nu=1)\,G_0\,v_{\pi}^{(0)} \nonumber\\
&&\!\!+\,V_1\,V_2\,V_3\,V_4\,\left[2\,\frac{\omega_1^2+\omega_2^2}
{\omega_1^3\,\omega_2^3}+
\frac{1}
{\omega_1^2\,\omega_2^2}\right] \nonumber\\
&&\!\!\times\left(E_i-\widetilde{E}_1-\widetilde{E}_2\right)\ ,
\end{eqnarray}
with $v_\pi^{(2)}(\nu=1)$ as given in Eq.~(\ref{eq:vpi2n}), provided the
relevant nucleon energies are considered.

\begin{figure*}[bthp]
\center{
\includegraphics[width=5in]{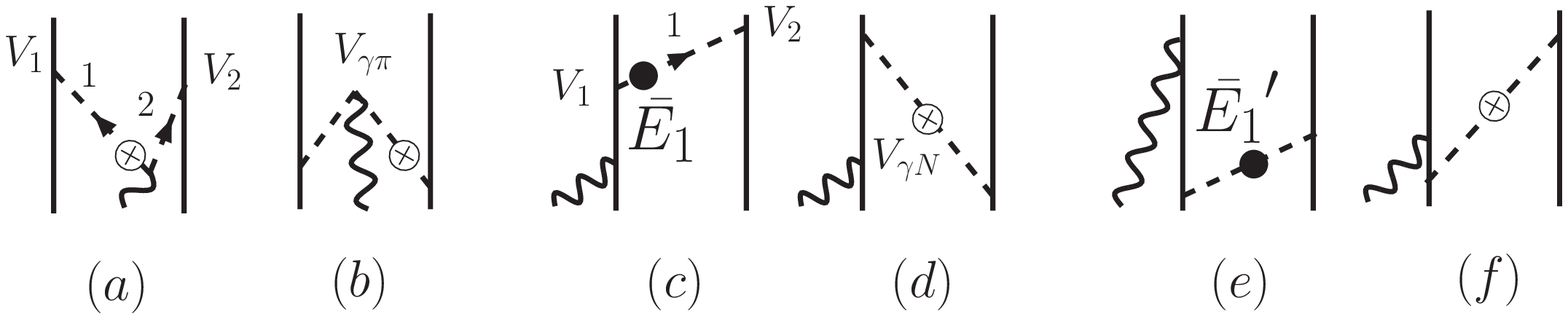}
}
\caption{Diagrams illustrating recoil-corrected OPE
amplitudes contributing to the charge operator at N3LO.
Only one among the possible time orderings is shown.
Notation is as in Figs.~\protect\ref{fig:fig1} and~\protect\ref{fig:fig5}.}
\label{fig:fig6}
\end{figure*}

To complete the evaluation of the box amplitude, we need
the expression of the irreducible contribution, which is given by
\begin{eqnarray}
 \label{eq:v2pi_irr}
T_{2\pi,\,\rm irr}^{(3)}&=&-\,
V_1\,V_2\,V_3\,V_4\,
\frac{E_i-\widetilde{E}_1-\widetilde{E}_2}
{\omega_1^2\,\omega_2^2}\nonumber\\
&&-2\,V_1\,V_4\,V_3\,V_2\bigg[
\frac{E_1-\widetilde{E}_1+E_2^{\, \prime}-\widetilde{E}^{\, \prime}_2}
{\omega_1^3\,\omega_2}\nonumber\\
&&+ \frac{E_1^{\, \prime}-\widetilde{E}_1+E_2-\widetilde{E}^{\, \prime}_2}
{\omega_1\,\omega_2^3} \bigg] \ ,
\end{eqnarray}
where, referring to Fig.~\ref{fig:fig5},
the first term is from the (irreducible) direct
diagrams of the type shown in panel (c), while the last two
are generated by the diagrams of the type shown in panel (d).
The sum of the reducible and irreducible pieces can then be written as
\begin{equation}
\label{eq:v2pi_final}
T_{2\pi}^{(3)}(\nu)
= v_{\pi}^{(0)}\,G_0\,v_{\pi}^{(2)}(\nu)
+ v_{\pi}^{(2)}(\nu)\,G_0\,v_{\pi}^{(0)} 
+v_{2 \pi}^{ {(3)}}(\nu)\ ,
\end{equation}
where $v_{2 \pi}^{(3)}(\nu)$ for $\nu=0,1$ are as given in
Eqs.~(\ref{eq:v3box}) and~(\ref{eq:v3nu}).

An analysis similar to that outlined above
leads to the following expressions for the ``triangle''
TPE, panel (b) of Fig.~\ref{fig:fig4}, and OPE contact,
panels (c) and (d), amplitudes and corresponding potentials:
\begin{eqnarray}
&&\!\!\!\!\!\!\!\!\!\!\!\!\!
T^{(3)}_{2\pi, \vartriangle}=v_{2\pi,\vartriangle}^{(3)}=
-\frac{g_A^2}{F_\pi^4}\, {\bm \tau}_1\cdot{\bm \tau}_2
\int \frac{1}{\omega_1^2\,\omega_2^2}
( {\bf q}_1\cdot{\bf q}_2\nonumber\\
&&\qquad\,\!\!\!\!\!\!\!\!\!+\,
i\, {\bm \sigma}_1\cdot{\bf q}_2\times {\bf q}_1)
(E_1+E_1^{\, \prime} -2 \, \widetilde{E}_1)
+1\rightleftharpoons2 , \\
&&\!\!\!\!\!\!\!\!\!\!\!
\!\!\!\!\!\!\!\!\!
T_{\rm CT,\, c}^{(3)}(\nu)=v_{\rm CT}^{(0)}\,G_0\,v_{\pi}^{(2)}(\nu)
+ v_{\pi}^{(2)}(\nu)\,G_0\,v_{\rm CT}^{(0)}\nonumber\\
&&\qquad\,\!\!\!\!\!\!\!\!
+\,v_{ \rm CT,\,c}^{(3)}(\nu)\ ,\\
&&\!\!\!\!\!\!\!\!\!\!\!\!\!
T_{\rm CT,\, d}^{(3)}=v_{ \rm CT,\, d}^{(3)}= \frac{3\,g_A^2}{2\,F_\pi^2}\,
\,\int
\frac{{\bm \sigma}_1\cdot{\bf q}_1\,v_{\rm CT}^{(0)}
\,{\bm \sigma}_1\cdot{\bf q}_1}{\omega_1^4}\nonumber\\
&&\qquad\,\!\!\!\!\!\!\!\!
\times
(E_1-\widetilde{E}_1+E_1^\prime-\widetilde{E}_1^\prime)+
1\rightleftharpoons2\ ,
\end{eqnarray}
where
\begin{equation}
\label{eq:vct0}
 v_{\rm CT}^{(0)}=C_S+C_T\,{\bm \sigma}_1\cdot{\bm \sigma}_2 \ ,
\end{equation}
is the contact potential at LO, while the N3LO potential
arising from the diagrams of panel (c) with the choices $\nu=0,1$
for the OPE potential $v_\pi^{(2)}(\nu)$, is given by
\begin{eqnarray}
 v_{ \rm CT,\,c}^{(3)}(\nu=0)\!\!&=&\!\! \frac{g_A^2}{2\,F_\pi^2}\,
{\bm \tau}_1\cdot{\bm \tau}_2\,\int
\frac{{\bm \sigma}_1\cdot{\bf q}_1\,v_{\rm CT}^{(0)}
\,{\bm \sigma}_2\cdot{\bf q}_1}{\omega_1^4}\nonumber\\
&\times&\!\!
(E_2-\widetilde{E}_2+E_1^\prime-\widetilde{E}_1^\prime)+
1\rightleftharpoons2\ ,\\
 v_{ \rm CT,\,c}^{(3)}(\nu=1)\!\!&=&\!\!
v_{ \rm CT,\,c}^{(3)}(\nu=0) +\!\!\int
\frac{v_{\rm CT}^{(0)}\,{v_{\pi}^{(0)}}}{2\,\omega_1^2}
(E_i-\widetilde{E}_1-\widetilde{E}_2)\nonumber\\
&+&
\int
\frac{{v_{\pi}^{(0)}}\,v_{\rm CT}^{(0)}}{2\,\omega_1^2}
(E_i-\widetilde{E}_1^\prime-\widetilde{E}_2^\prime)\ ,
\end{eqnarray}
and the nucleon energies and
pion momenta are defined in Fig.~\ref{fig:fig4}.

\subsection{OPE Charge Operators at N3LO}
\label{app:rc_rho_ope}

Before turning our attention to the N4LO
corrections, we outline the derivation of
the N3LO OPE charge operators whose expressions
have been given in Eqs.~(\ref{eq:rho_gpipi})--(\ref{eq:rhon1}).
As discussed in Sec.~\ref{sec:cnt3}, these
operators vanish in the static limit (that is, at
N2LO), while at N3LO they are given by amplitudes
associated with the diagrams of the type illustrated in Fig.~\ref{fig:fig6}.
In particular, the charge operator in Eq.~(\ref{eq:rho_gpipi}),
which we denote as $\rho^{(0)}_{\gamma \pi}$ for
later convenience, is obtained as
\begin{equation}
\label{eq:rho_pi_pi_g}
\rho_{\gamma \pi}^{(0)}=
-\,4 \frac{V_1\,V_2\,V_{\gamma\pi}}
{\omega_1\,\omega_2}\,(E_1-E^{\, \prime}_1-E_2+E^{\, \prime}_2) \ .
\end{equation}
The vertex $V_{\gamma\pi}$ is proportional to that
associated with the interaction Hamiltonian $H_{\gamma\pi}$
in Sec.~\ref{sec:hint},
\begin{equation}
 -i\,e\,\, \epsilon_{abz}\,\,\frac{\omega_1-\omega_2}
{\sqrt{4\,\omega_1\,\omega_2}} \  ,
\end{equation}
where the pion energies are as indicated in
panel (a) of Fig.~\ref{fig:fig6}.  It is convenient to factor
out the energy numerator $(\omega_1-\omega_2)$, and define
$V_{\gamma\pi}$ as
\begin{equation}
V_{\gamma\pi}=-i\,e\, \epsilon_{abz}\,\frac{1}{\sqrt{4\,\omega_1\,\omega_2}} \  .
\end{equation}
Next, we consider the amplitude associated with the diagrams shown in panels
(c)--(f) of Fig.~\ref{fig:fig6}:
\begin{eqnarray}
T^{(0)}_{\gamma {\rm d}}(\nu)&=&\left[v_\pi^{(2)}(\nu)\,G_0\,\rho^{(-3)}
+\rho^{(-3)}\,G_0\,v_\pi^{(2)}(\nu)\right]  \nonumber \\
&&+\rho_{\rm d}^{(0)}(\nu) \ ,
\end{eqnarray}
where $\rho^{(-3)}$ is the LO charge operator given in
Eq.~(\ref{eq:r-3}), while $\rho_{\rm d}^{(0)}(\nu)$ is the
N3LO OPE contribution defined in
Eqs.~(\ref{eq:rhon0})--(\ref{eq:rhon1})
for $\nu=0,1$.  The latter is written as
\begin{equation}
\label{eq:rho-d-ope}
\rho_{\rm d}^{(0)}(\nu)=\rho_{\gamma N}^{(0)}(\nu)+\rho_{N\gamma}^{(0)}(\nu)
+1\rightleftharpoons2\ ,
\end{equation}
where $\rho_{\gamma N}^{(0)}$ comes from
the diagrams shown in panels (c) and (d) of Fig.~\ref{fig:fig6},
while $\rho_{N\gamma}^{(0)}$ is associated with those
of panels (e) and (f). For $\nu=0$, we find
\begin{eqnarray}
 \label{eq:rho_g_N_pi}
&&\!\!\!\!\!\!\!\!\!\!\!\!\!\!\!\!\!\rho_{\gamma N}^{(0)}(\nu=0)=
\frac{V_1\,V_2\,V_{\gamma N}}
{\omega_1^3}\,(E_1^{\, \prime} -\overline{E}_1+E_2-E^{\, \prime}_2)\ ,\\
 \label{eq:rho_pi_N_g}
&&\!\!\!\!\!\!\!\!\!\!\!\!\!\!\!\!\!\rho_{ N\gamma}^{(0)}(\nu=0)=-\frac{V_{N\gamma}\,V_1\,V_2}{\omega_1^3}
(\overline{E}_1^{\, \prime} -E_1+E_2-E^{\, \prime}_2)\ ,
\end{eqnarray}
while for $\nu=1$ we obtain
\begin{eqnarray}
 \rho_{\gamma N}^{(0)}(\nu=1)&=&
2\,\frac{V_1\,V_2\,V_{\gamma N}}
{\omega_1^3}\,(E_2-E^{\, \prime}_2) \ , \\
\rho_{ N\gamma}^{(0)}(\nu=1)&=&-2\,\frac{V_{\gamma N}\,V_1\,V_2} 
{\omega_1^3}\,(E_2-E^{\, \prime}_2)\ .
\end{eqnarray}
The energies $\overline{E}_1$ and $\overline{E}_1^{\, \prime}$ are as indicated in panels (c)
and (e) of Fig.~\ref{fig:fig6}, respectively, and
$V_{\gamma N}=e\,e_{N,1}$ is the vertex implied by the interaction
Hamiltonian $H_{\gamma N}$ at LO.

\subsection{N4LO Charge Operators}
\label{app:n4lo_energy}

\begin{figure}[tp]
\centerline{
\includegraphics[width=3.3in]{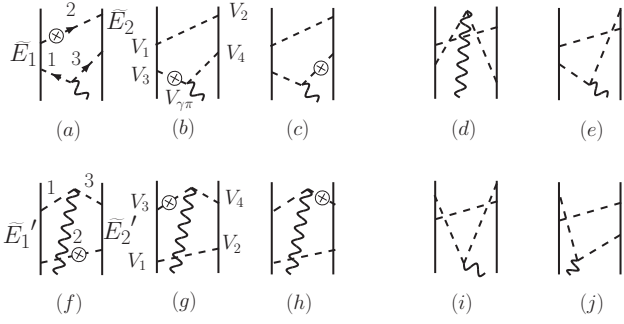}
}
\caption{Diagrams illustrating the static, panels (d), (e), (i), and (j), and
recoil-corrected, remaining panels, diagrams associated with the N4LO contribution
shown in panel (c) of Fig.~\ref{fig:fig2}.
Only one among the possible time orderings is shown.
Notation is as in Figs.~\protect\ref{fig:fig1} and~\protect\ref{fig:fig5}.}
\label{fig:fig7}
\end{figure}
We can now proceed to sketch the derivation of the
charge operators illustrated in panels (c)--(j)
of Fig.~\ref{fig:fig2}.  The one-loop corrections of panels (c)--(e)
involve a ${\gamma \pi \pi}$ electromagnetic vertex,
while a $\gamma N N$ interaction enters those of panels
(f)--(j).  We give details on the derivation
of the operators $\rho^{(1)}_{\rm c}$ and $\rho^{(1)}_{\rm h}$
as representatives of these classes of diagrams.

First, we consider the pion-in-flight term.
The diagrams contributing at N4LO are shown in
Fig.~\ref{fig:fig7}.   As diagrammatically shown in the figure, the
irreducible contributions, panels (d), (e), (i), and (j),
are evaluated in the static limit,
while next-to-leading order terms in the expansion
of energy denominators are retained in
the evaluation of the reducible contributions.
We write the total amplitude as a sum of these,
that is
\begin{eqnarray}
\label{eq:T_c}
 T_{\gamma {\rm c}}^{(1)}
&=&T_{\gamma{\rm c,\,red}}^{(1)}
+T_{\gamma{\rm c,\,irr}}^{(1)} \ .
\end{eqnarray}
The reducible and irreducible contributions are given by
\begin{eqnarray}
\label{eq:rho_c_red}
\!\!\!\!\!\!\!\!\!\!T_{\gamma{\rm c,\,red}}^{(1)}\!\!&=& \!\!  v_\pi^{(0)}\,G_0\,
\rho_{\gamma\pi}^{(0)}+
\rho_{\gamma\pi}^{(0)}\,G_0\,v_\pi^{(0)}\nonumber\\
&+& \!\! \frac{2\,(\omega_1-\omega_3)}
{\omega_1\,\omega_2\,\omega_3\,(\omega_1+\omega_3)}\,
\left[V_1\,V_2\, ,\,V_3\,V_4\right]V_{\gamma\pi} \ , \\
\label{eq:rho_c_irr}
\!\!\!\!\!\!\!\!\!\!T_{\gamma{\rm c, \, irr}}^{(1)}\!\!&=& \!\!-\frac{2\,(\omega_1-\omega_3)}
{\omega_1\,\omega_2\,\omega_3\,(\omega_1+\omega_3)}
\left[V_1\,V_2\, ,\,V_3\,V_4\right]V_{\gamma\pi}\nonumber\\
 &-&\!\!\frac{8}{\omega_1\,\omega_2\,\omega_3}\,
\left[V_1\,V_4\, ,\,V_3\,V_2\right]V_{\gamma\pi} \ ,
\end{eqnarray}
where $[\dots,\dots]$ denotes a commutator and
$\rho_{\gamma \pi}^{(0)}$ is the N3LO OPE charge
operator defined in Eq.~(\ref{eq:rho_pi_pi_g}).
In Eq.~(\ref{eq:rho_c_red}), the terms multiplied by the spin-isospin
combinations $ V_1\,V_2\,V_3\,V_4\,V_{\gamma\pi}$ and
$ V_3\,V_4\,V_1\,V_2 \,V_{\gamma\pi}$ 
come from the diagrams shown in panels (a)--(c)
and (f)--(h), respectively, whereas in Eq.~(\ref{eq:rho_c_irr})
the first term results from the evaluation of the diagrams
shown in panels (d) and (i) of Fig.~\ref{fig:fig7} and the second one is
from those illustrated in panels (e) and (j).
The total amplitude is then given by
\begin{eqnarray}
 T_{\gamma{\rm c}}^{(1)}
&=&v_\pi^{(0)}\,G_0\,
\rho_{\gamma\pi}^{(0)}+
\rho_{\gamma\pi}^{(0)}\,G_0\,v_\pi^{(0)}
+\rho_{\rm c}^{(1)} \ ,
\end{eqnarray}
where
\begin{equation}
\rho^{(1)}_{\rm c}= -\frac{8}{\omega_1\,\omega_2\,\omega_3}\,
\left[V_1\,V_4\, ,\,V_3\,V_2\right]\,V_{\gamma\pi} \ ,
\end{equation}
which, after resolving the spin-isospin structure implied
by the vertices, reduces to the N4LO charge operator
in Eq.~(\ref{eq:rho_c}).

\begin{figure}[tp]
\centerline{
\includegraphics[width=3.3in]{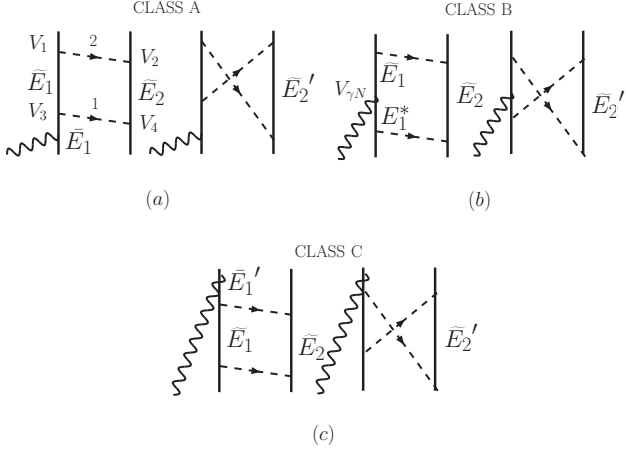}
}
\caption{Classes of diagrams associated with the
one-loop contribution illustrated in panel (h) of
Fig.~\ref{fig:fig2}. Notation is as in Fig.~\protect\ref{fig:fig1}.}
\label{fig:fig8}
\end{figure}

We now turn our attention to the one-loop correction
shown in panel (h) of Fig.~\ref{fig:fig2}.
For this contribution, we distinguish among
three classes of diagrams depending on whether
the photon is absorbed before pion
one, class A, after pion one, class B, or after pion
two, class C.  These classes are represented in Fig.~\ref{fig:fig8}---the
vertices and kinetic energies of intermediate nucleons are
as indicated in the figure.

\begin{figure*}[tp]
\centerline{
\includegraphics[width=4.6in]{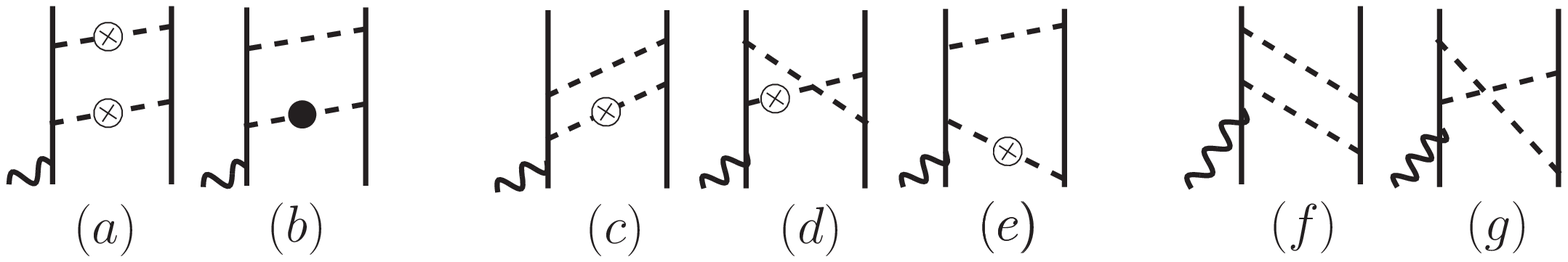}
}
\caption{
Diagrams illustrating the static, panels (f) and (g), and
recoil-corrected, remaining panels, class A diagrams contributing
at N4LO. In panels (b)--(e),
the crossed and full circles can be either on pion one
(as shown in the figure) or on pion two.
Only one among the possible time orderings is shown.
Notation is as in Figs.~\protect\ref{fig:fig1} and~\protect\ref{fig:fig5}.
}
\label{fig:fig9}
\end{figure*}

We start off by discussing the result obtained
for the class A amplitude.  In Fig.~\ref{fig:fig9} we show
the diagrams contributing at N4LO.  The
irreducible diagrams, panels (f) and (g) of this figure, are
evaluated in the static limit.  The N4LO recoil-corrected
contributions associated with the single (double) reducible
diagrams, panels (c)--(e) of Fig.~\ref{fig:fig9} [panels (a) and (b)],
are obtained by retaining $Q^{\,0}$ ($Q^{\,1}$) terms in the
expansions of energy denominators involving pions.
The N4LO amplitude is then written as
\begin{eqnarray}
 T_{\gamma{\rm A}}^{(1)}(\nu)
&=&T_{\gamma{\rm A,\,red}}^{(1)}(\nu) +T_{\gamma{\rm A,\,irr}} \ ,
\end{eqnarray}
where
\begin{eqnarray}
\label{eq:rho_gredA}
T_{\gamma{\rm A,\,red}}^{(1)}(\nu=0)\!\!&=& \!\!\bigg[
v_{\pi}^{(0)}\,G_0\,v_{\pi}^{(2)}(\nu=0)\,G_0\,
\rho^{(-3)}\nonumber\\
&&\!\!\!\!\!\!\!\!\!\!
\!\!\!\!\!\!\!\!\!\!\!\!\!\!\!\!
+ \,v_{\pi}^{(2)}(\nu=0)\,G_0\,v_{\pi}^{(0)}\,G_0\,
\rho^{(-3)}\nonumber\\
&&\!\!\!\!\!\!\!\!\!\!
\!\!\!\!\!\!\!\!\!\!\!\!\!\!\!\!
+\,v_{2 \pi}^{ {(3)}}(\nu=0)\,G_0\,
\rho^{(-3)}+v_{\pi}^{(0)}\,G_0\,\rho_{\gamma N}^{(0)}(\nu=0)\bigg] \nonumber\\
&&\!\!\!\!\!\!\!\!\!\!
\!\!\!\!\!\!\!\!\!\!\!\!\!\!\!\!
+\,\bigg[\frac{2}{\omega_1^3\,\omega_2}-
\frac{1}{\omega_1\,\omega_2(\omega_1+\omega_2)^2}-
\frac{1}{\omega_1^2\,\omega_2(\omega_1+\omega_2)}\bigg]\nonumber\\
&&\!\!\!\!\!\!\!\!\!\!
\!\!\!\!\!\!\!\!\!\!\!\!\!\!\!\!
\times \,V_1\,V_2\,V_3\,V_4\,V_{\gamma N}
\nonumber\\
&&\!\!\!\!\!\!\!\!\!\!
\!\!\!\!\!\!\!\!\!\!\!\!\!\!\!\!
+
\bigg[\frac{1}{\omega_1\,\omega_2\,(\omega_1+\omega_2)^2}
-\frac{2}{\omega_1^3\,\omega_2}
-\frac{2}{\omega_1\,\omega_2^3}\bigg]\nonumber\\
&&\!\!\!\!\!\!\!\!\!\!
\!\!\!\!\!\!\!\!\!\!\!\!\!\!\!\!
\times 
\, V_1\,V_4\,V_3\,V_2\,V_{\gamma N}
+1\rightleftharpoons2 \ .
\end{eqnarray}
The N4LO LS terms arising from the reducible diagrams are
listed in the first two lines of the equation above,
where $v_{\pi}^{(0)}$, $v_{\pi}^{(2)}(\nu=0)$ and
$v_{2\pi}^{(3)}(\nu=0)$ are the LO, N2LO and N3LO components
of the $NN$ potential given in Eqs.~(\ref{eq:vpi0}),~(\ref{eq:vpi2}),
and~(\ref{eq:v3box}), respectively, while the $\rho^{(-3)}$
and $\rho_{\gamma N}^{(0)}$ charge operators have been defined in
Eqs.~(\ref{eq:r-3}) and~(\ref{eq:rho_g_N_pi}).
The last two terms in Eq.~(\ref{eq:rho_gredA}) constitute
the N4LO recoil-corrected contribution associated with the reducible diagrams.
In particular, the second term is generated by the direct
reducible diagrams of panels (a)--(c)
and (e), while the last one is obtained from the contributions
of type (d).

The irreducible amplitude from diagrams
in panels (f) and (g) of Fig.~\ref{fig:fig9} reads
\begin{eqnarray}
\label{eq:rho_girrA}
T_{\gamma{\rm A,\,irr}}^{(1)}&=& 
\left[
\frac{1}{\omega_1\,\omega_2(\omega_1+\omega_2)^2}+
\frac{1}{\omega_1^2\,\omega_2\,(\omega_1+\omega_2)}\right]\nonumber\\
&\times&
V_1\,V_2\,V_3\,V_4\,V_{\gamma N}\nonumber\\
&+&\left[\frac{2}{\omega_1\,\omega_2^3}-
\frac{1}{\omega_1\,\omega_2(\omega_1+\omega_2)^2}
\right]\nonumber\\
&\times&V_1\,V_4\,V_3\,V_2\,V_{\gamma N}
+1\rightleftharpoons2 \ ,
\end{eqnarray}
and combining Eqs.~(\ref{eq:rho_gredA}) and~(\ref{eq:rho_girrA})
leads to a total amplitude, which can be written as
\begin{eqnarray}
\label{eq:T-A}
T_{\gamma{\rm A}}^{(1)}(\nu)
&=&
v_{\pi}^{(0)}\,G_0\,v_{\pi}^{(2)}(\nu)\,G_0\,
\rho^{(-3)}\nonumber\\
&+& v_{\pi}^{(2)}(\nu)\,G_0\,v_{\pi}^{(0)}\,G_0\,
\rho^{(-3)}
+v_{2 \pi}^{ {(3)}}(\nu)\,G_0\,
\rho^{(-3)}\nonumber\\
&+& v_{\pi}^{(0)}\,G_0\,\rho_{\gamma N}^{(0)}(\nu)
+\rho_{\rm A}^{(1)}(\nu) +1\rightleftharpoons2\ ,
\end{eqnarray}
where 
\begin{equation}
\label{eq:rho_A}
 \rho_{\rm A}^{(1)}(\nu=0) =\frac{2}{\omega_1^3\,\omega_2}\,
V_1\,V_3\,\left[V_2\, ,\,V_4\right]\,V_{\gamma N}\ .
\end{equation}

The derivation of the class C amplitude (see
Fig.~\ref{fig:fig8}) is analogous to that described
above.  We find:
\begin{eqnarray}
\label{eq:T-C}
\!\!\!\!\!\!\!\!T_{\gamma {\rm C}}^{(1)}(\nu)
&=&
\rho^{(-3)}\,G_0\,v_{\pi}^{(0)}\,G_0\,v_{\pi}^{(2)}(\nu)\nonumber\\
&+& \rho^{(-3)}\,G_0\,v_{\pi}^{(2)}(\nu)\,G_0\,v_{\pi}^{(0)} +
\rho^{(-3)}\,G_0\,v_{2 \pi}^{ {(3)}}(\nu)\nonumber\\
&+&\rho_{N\gamma }^{(0)}(\nu)\,G_0\,v_{\pi}^{(0)}+\rho_{\rm C}^{(1)}(\nu)
+1\rightleftharpoons2\ ,
\end{eqnarray}
where $\rho_{N\gamma }^{(0)}$ is the OPE charge operator
defined in Eq.~(\ref{eq:rho_pi_N_g}), and
\begin{eqnarray}
\label{eq:rho_C}
 \rho_{\rm C}^{(1)}(\nu=0) &=&\frac{2}{\omega_1\,\omega_2^3}\,
\,V_{\gamma N} V_1\,V_3\,\left[V_2\, ,\,V_4\right]
\ .
\end{eqnarray}

\begin{figure*}[tp]
\centerline{
\includegraphics[width=4.6in]{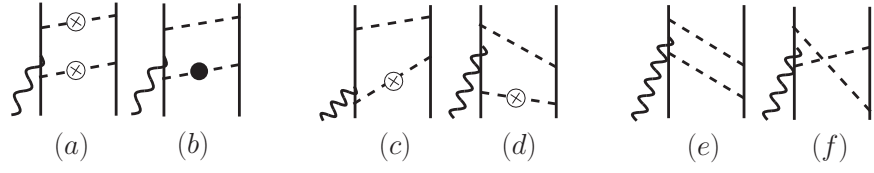}
}
\caption{
Diagrams illustrating the static, panels (e) and (f), and
recoil-corrected, remaining panels, class B diagrams contributing
at N4LO. In panels (b)--(d),
the crossed and full circles can be either on pion one
(as shown in the figure) or on pion two.
Only one among the possible time orderings is shown.
Notation is as in Figs.~\protect\ref{fig:fig1}
and~\protect\ref{fig:fig5}.
}
\label{fig:fig10}
\end{figure*}

Class B of box diagrams at N4LO are shown in
Fig.~\ref{fig:fig10}.  The reducible amplitude, associated with the diagrams
in panels (a)--(d), is found to be
\begin{eqnarray}
\label{eq:rho_gredB}
T_{\,\gamma{\rm B,\, red}}^{(1)}(\nu=0)\!\!\!&=& \!\!
v_{\pi}^{(0)}\,G_0\,\rho^{(-3)}\,G_0\,v_{\pi}^{(2)}(\nu=0)\nonumber\\
&&\!\!\!\!\!\!\!\!\!\!
\!\!\!\!\!\!\!\!\!\!\!\!\!\!\!\!\!\!\!\!\!\!\!\!\!\!\!\!\!\!\!
+\,
v_{\pi}^{(2)}(\nu=0)\,G_0\,\rho^{(-3)}\,G_0\,v_{\pi}^{(0)}\nonumber\\
&&\!\!\!\!\!\!\!\!\!\!
\!\!\!\!\!\!\!\!\!\!\!\!\!\!\!\!\!\!\!\!\!\!\!\!\!\!\!\!\!\!\!
+
\rho_{\gamma N }^{(0)}(\nu=0)\,G_0\,v_{\pi}^{(0)}+
v_{\pi}^{(0)}\,G_0\,\rho_{ N \gamma}^{(0)}(\nu=0)\\
&&\!\!\!\!\!\!\!\!\!\!
\!\!\!\!\!\!\!\!\!\!\!\!\!\!\!\!\!\!\!\!\!\!\!\!\!\!\!\!\!\!\!
-\bigg(2\,\frac{\omega_1^2+\omega_2^2}{\omega_1^3\,\omega_2^3}
+\frac{1}{\omega_1^2\,\omega_2^2}\bigg)
V_1V_2V_{\gamma N}V_3V_4
\!+\!1\rightleftharpoons2  \ , \nonumber
\end{eqnarray}
while the irreducible amplitude, corresponding to the diagrams
in panels (e) and (f), amounts to
\begin{eqnarray}
\label{eq:rho_girrB}
T_{\gamma{\rm B,\,irr}}^{(1)}\!\!&=& \!\!
\frac{1}{\omega_1^2\,\omega_2^2}\,V_1\,V_2\,V_{\gamma N}\,V_3\,V_4\nonumber\\
&+ &\!\!2\,\frac{\omega_1^2+\omega_2^2}{\omega_1^3\,\omega_2^3}
\,V_1\,V_4\,V_{\gamma N}\,V_3\,V_2
+1\rightleftharpoons2 \ ,
\end{eqnarray}
and the class B amplitude is then given by
\begin{eqnarray}
\label{eq:T-B}
T_{\gamma{\rm B}}^{(1)}(\nu)&=&
 v_{\pi}^{(0)}\,G_0\,\rho^{(-3)}\,G_0\,v_{\pi}^{(2)}(\nu)
\nonumber\\
&+& v_{\pi}^{(2)}(\nu)\,G_0\,\rho^{(-3)}\,G_0\,v_{\pi}^{(0)}\nonumber\\
&+&\rho_{\gamma N }^{(0)}(\nu)\,G_0\,v_{\pi}^{(0)}+
v_{\pi}^{(0)}\,G_0\,\rho_{ N \gamma}^{(0)}(\nu)\nonumber\\
&+&\rho_{\rm B}^{(1)}(\nu) 
+1\rightleftharpoons2 \ ,
\end{eqnarray}
where 
\begin{equation}
\label{eq:rho_B}
 \rho_{\rm B}^{(1)}(\nu=0) =\,-\,2 \,\frac{\omega_1^2+\omega_2^2}
{\omega_1^3\,\omega_2^3}\,\,V_1\,V_{\gamma N}
\,V_3\,\left[V_4\, ,\,V_2\right]\ .
\end{equation}

Finally, the total amplitude associated with the diagram shown
in panel (h) of Fig.~\ref{fig:fig2} is given by the sum of the
A, B, and C amplitudes, that is
\begin{eqnarray}
 T_{\rm h}^{(1)}(\nu)&=&T_{\gamma{\rm A}}^{(1)}(\nu)+T_{\gamma{\rm B}}^{(1)}(\nu)
+T_{\gamma{\rm C}}^{(1)}(\nu)
\nonumber\\
&=&\bigg[\left[ \rho^{(-3)}\,G_0\,v_{\pi}^{(0)}\,G_0\,v_{\pi}^{(2)}(\nu)
+{\rm permutations}\right]\nonumber\\
&+&
\left[ \rho^{(-3)}\,G_0\,v_{2 \pi}^{ {(3)}}(\nu)
+v_{2 \pi}^{ {(3)}}(\nu)\,G_0\,\rho^{(-3)} \right]
+1\rightleftharpoons2\bigg]
\nonumber\\
&+&\left[\rho_{\rm d}^{(0)}(\nu)\,G_0\,v_{\pi}^{(0)}
+v_{\pi}^{(0)}\,G_0\,\rho_{\rm d}^{(0)}(\nu)\right]
\nonumber\\
& +& \rho_{\rm h}^{(1)}(\nu) ,
\end{eqnarray}
where $\rho_{\rm d}^{(0)}$ is defined as in Eq.~(\ref{eq:rho-d-ope}),
while
\begin{eqnarray}
\rho_{\rm h}^{(1)}(\nu)= \rho_{\rm A}^{(1)}(\nu)
+\rho_{\rm B}^{(1)}(\nu)+\rho_{\rm C}^{(1)}(\nu) +1\rightleftharpoons2 \ .
\end{eqnarray}
For $\nu=0$, $\rho_{\rm A}^{(1)}$, $\rho_{\rm B}^{(1)}$, and
$\rho_{\rm C}^{(1)}$ are as given in Eqs.~(\ref{eq:rho_A}),
~(\ref{eq:rho_B}), and~(\ref{eq:rho_C}), respectively, and
carrying out the spin-isospin algebra leads to Eq.~(\ref{eq:rho_g}).
Similarly, for $\nu=1$ we find:
\begin{eqnarray}
 &&\!\!\!\!\!\!\!\!\!\!\!\!\!\!\!\!\!\!\!\!\!\!
\rho_{\rm A}^{(1)}(\nu=1)=\rho_{\rm A}^{(1)}(\nu=0)+
\frac{2}{\omega_1\,\omega_2^3}
\,V_1\,V_2\,V_3\,V_4\,V_{\gamma N}\ , \\
 &&\!\!\!\!\!\!\!\!\!\!\!\!\!\!\!\!\!\!\!\!\!\!
\rho_{\rm B}^{(1)}(\nu=1)=\rho_{\rm B}^{(1)}(\nu=0) \ , \\
 &&\!\!\!\!\!\!\!\!\!\!\!\!\!\!\!\!\!\!\!\!\!\!
\rho_{\rm C}^{(1)}(\nu=1)=\rho_{\rm C}^{(1)}(\nu=0)
-\frac{2}{\omega_1\,\omega_2^3}\,V_{\gamma N}\,V_1\,V_2\,V_3\,V_4\ ,
\end{eqnarray}
from which it follows that
\begin{eqnarray}
\hspace*{-0.33in}\rho_{\rm h}(\nu=1)\!\!&=&\!\! \rho_{\rm h}(\nu=0)\nonumber\\
\!\!&+&\!\!\left[  \frac{2}{\omega_1\,\omega_2^3}
\,[V_1\,V_2\,V_3\,V_4\,,\,V_{\gamma N}]+1\rightleftharpoons2   \right] \ ,
\end{eqnarray}
and simplifying the spin-isospin structures
leads to the operator $\rho^{(1)}_{\rm h}(\nu=1)$ given in Eq.~(\ref{eq:rho_gn1}).

Below we list the expressions for the amplitudes
associated with the remaining N4LO one-loop
corrections illustrated in Fig.~\ref{fig:fig2},
in particular, referring to panels (d), (e), (g),
(i), and (j) of this figure (type (f) operator vanishes
as pointed out in Sec.~\ref{sec:cnt4}) we obtain:
\begin{eqnarray}
\!\!\!\!\!\!\!\!\!
\!\!\!\!\!\!\!\!\!
T_{\rm d}^{(1)}\!\!\!&=&\!\!\!
\left[v_{\rm CT}^{(0)}\,G_0\,\rho_{\gamma \pi}^{(0)}
+ \rho_{\gamma \pi}^{(0)}\,G_0\,v_{\rm CT}^{(0)}\right]
\ ,\\
T_{\rm e}^{(1)}\!\!\!&=&\!\!\!\rho_{\rm e}^{(1)} \ , \\
T_{\rm g}^{(1)}\!\!\!&=&\!\!\!
\left[v_{2\pi, \vartriangle}^{(3)}\,G_0\,\rho^{(-3)}
+ \rho^{(-3)}\,G_0\,v_{2\pi, \vartriangle}^{(3)}+1
\rightleftharpoons2\right]\nonumber\\
&+&\!\!\!\rho_{\rm g}^{(1)} \ ,\\
T_{\rm i}^{(1)}(\nu)\!\!\!&=&\!\!\!\bigg[\!
\left[ \rho^{(-3)}\,G_0\,v_{\rm CT}^{(0)}\,G_0\,v_{\pi}^{(2)}(\nu)
+{\rm permutations}\right]\nonumber\\
&+&\!\!\!
\big[ \rho^{(-3)}\,G_0\,v_{\rm CT, \,c}^{ {(3)}}(\nu)+
v_{\rm CT, \,c}^{ {(3)}}(\nu)\,G_0\,\rho^{(-3)} \big]
+1\rightleftharpoons2\bigg]
\nonumber\\
&+&\!\!\!
\left[\rho_{\rm d}^{(0)}(\nu)\,G_0\,v_{\rm CT}^{(0)}
+v_{\rm CT}^{(0)}\,G_0\,\rho_{\rm d}^{(0)}(\nu)\right]\nonumber\\
&+&\!\!\!\rho_{\rm i}^{(1)}(\nu)\ ,\\
T_{\rm j}^{(1)}\!\!\!&=&\!\!\!\left[v_{\rm CT, \,d}^{(3)}\,G_0\,\rho^{(-3)}
+ \rho^{(-3)}\,G_0\,v_{\rm CT, \,d}^{(3)}+1\rightleftharpoons2\right]
\nonumber\\
&+&\!\!\!\rho_{\rm j}^{(1)} \ ,
\end{eqnarray}
where $\rho_{\rm e}^{(1)}$, $\rho_{\rm g}^{(1)}$, and $\rho_{\rm j}^{(1)}$
are given in Eqs.~(\ref{eq:rho_e}), ~(\ref{eq:rho_f}), and~(\ref{eq:rho_i}),
respectively, while $\rho_{\rm i}^{(1)}$ with $\nu=0, 1$ vanishes.
In the equations above the LS terms involve the LO contact,
Eq.~(\ref{eq:vct0}), and N2LO OPE components of the $NN$
potential, Eqs.~(\ref{eq:vpi2}) and ~(\ref{eq:vpi2n}),
as well as the N3LO potential $v^{(3)}$, derived in Sec.~\ref{app:rc_NNpot}.
In addition, the $\rho^{(-3)}$ is defined in Eq.~(\ref{eq:r-3}),
while the N3LO OPE charge operators, $\rho^{(0)}$,
are listed in Sec.~\ref{app:rc_rho_ope}.

\section{The N4LO charge operators in $r$-space}
\label{app:rspace}

The configuration-space representations of the charge operators
at order $n=-3, \dots, 0$ are well known~\cite{Schiavilla90}.  Here
we obtain those corresponding to $\rho^{(1)}$ for $\nu=0$ only.  Note that
$\rho^{(1)}_{\rm a}$, $\rho^{(1)}_{\rm d}$, and $\rho^{(1)}_{\rm i}(\nu=0)$ vanish.
Next, consider
\begin{eqnarray}
\label{eq:rho_br}
\rho^{(1)}_{\rm b} ({\bf q})\!\!&=\!\!&\!\! e\,\frac{2\,g_A^2}{F_\pi^4}\, \tau_{2,z}\!\!
\int_{{\bf k}_1,{\bf k}_2}\!\!{\rm e}^{i{\bf k}_1\cdot {\bf r}_1}\, {\rm e}^{i{\bf k}_2\cdot {\bf r}_2}\,
\overline{\delta}({\bf k}_1\!+\!{\bf k}_2\!-{\bf q}) \nonumber\\
&\times&
\int_{{\bf q}_1,{\bf q}_2}\!\! \frac{{\bf q}_1\cdot{\bf q}_2}
{\omega_1^2\,\omega_2^2}\, \overline{\delta}({\bf q}_1\!+\!{\bf q}_2\!-\!{\bf k}_1)\,
+1 \rightleftharpoons 2 \nonumber \\
&=&\!\!-e\,\frac{2\,g_A^2}{F_\pi^4}\, \tau_{2,z}\, {\rm e}^{i{\bf q}\cdot {\bf r}_2}
\left[{\bm \nabla} f_\pi(r) \right]\cdot
\left[{\bm \nabla} f_\pi(r) \right] \nonumber\\
&+&1 \rightleftharpoons 2 \ ,
\end{eqnarray}
where in the second line ${\bf r}$ denotes the relative position
${\bf r}={\bf r}_1-{\bf r}_2$ of the two nucleons, and
\begin{equation}
f_\pi(r) = \int_{\bf p} {\rm e}^{i{\bf p}\cdot {\bf r}} \, 
\frac{1}{p^2+m_\pi^2}=\frac{1}{4\pi}\frac{{\rm e}^{-m_\pi r}}{r} \ .
\end{equation}
Of course, the expression above is ill-behaved in the limit
of vanishing internucleon separations, and needs to be regularized.  This can be
accomplished by replacing
\begin{equation}
f_\pi(r) \rightarrow f_\Lambda(r) =
\int_{\bf p} {\rm e}^{i{\bf p}\cdot {\bf r}} \, 
\frac{C_\Lambda(p)}{p^2+m_\pi^2} \ ,
\end{equation}
and in applications so far~\cite{Girlanda10} the cutoff function has been taken as
$C_\Lambda(p)={\rm exp}(-p^4/\Lambda^4)$.  Similarly, we find:
\begin{eqnarray}
\label{eq:cr}
\rho^{(1)}_{\rm c}\!\!&=&\!\!\!- e\,\frac{2\,g_A^4}{F_\pi^4}\Big[ 2\, (\tau_{1,z}+\tau_{2,z})\nonumber\\
&\times&\!\!\!\big[
\epsilon_{\alpha \beta \gamma} \epsilon_{\lambda \mu \nu} \, \sigma_{1,\alpha} \,
 \sigma_{2,\lambda}+\delta_{\beta \gamma} \delta_{\mu \nu} \big]\nonumber\\
&-&\!\!\!({\bm \tau}_1\times{\bm \tau}_2)_z\,\big[ \delta_{\beta \gamma}\,\epsilon_{ \lambda \mu \nu }\, \sigma_{2,\lambda}-\epsilon_{\alpha \beta \gamma} \, \delta_{\mu \nu}\, \sigma_{1,\alpha} \big] \Big] \nonumber\\
&\times&\!\!\!
\left[ \partial_{1,\beta} \partial_{2,\mu} \, {\rm e}^{i {\bf q}\cdot{\bf R}} h_\pi({\bf r}) \right]
\,\left[\partial_\gamma \partial_\nu f_\pi(r)\right] \ ,\\
\label{eq:er}
\rho^{(1)}_{\rm e}\!\!&=&\!\!\!-e\,\frac{2\, g_A^2}{F_{\pi}^2}\,(\tau_{1,z}+\tau_{2,z})\,
{\rm e}^{i{\bf q}\cdot {\bf R}}\, \delta({\bf r})\, I({\bf q})  \\
\label{eq:fr}
\rho^{(1)}_{\rm g}\!\!&=&\!\!\! e\,\frac{2\,g_A^2}{F_\pi^4}\, \tau_{2,z}\, {\rm e}^{i{\bf q}\cdot {\bf r}_1}
\left[{\bm \nabla} f_\pi(r) \right]\cdot
\left[{\bm \nabla} f_\pi(r) \right] \nonumber\\
&+&1 \rightleftharpoons 2 \ , \\
\label{eq:gr}
\rho^{(1)}_{\rm h}(\nu=0)\!\!&=&\!\!\!
-e\,\frac{2\,g_A^4}{F_\pi^4}\, {\rm e}^{i{\bf q}\cdot {\bf R}}\Big[ 2\, \tau_{1,z}\,
\epsilon_{\alpha \beta \gamma} \epsilon_{\lambda \mu \nu} \, \sigma_{1,\alpha}\,
 \sigma_{2,\lambda}\, \nonumber\\
&+&\!\!\!2\, \tau_{2,z}\, \delta_{\beta \gamma} \delta_{\mu \nu}
-({\bm \tau}_1\times{\bm \tau}_2)_z\, \epsilon_{\alpha \beta \gamma } \, \delta_{\mu \nu}\, \sigma_{1,\alpha} \Big]\nonumber\\
&\times&
 \left[\partial_\beta \partial_\mu \tilde{f}_\pi(r)\right]\,
\left[\partial_\gamma \partial_\nu f_\pi(r)\right] +1 \rightleftharpoons 2 \ ,\\
\label{eq:ir}
\rho^{(1)}_{\rm j}\!\!&=&\!\!\! e\,\frac{2\, g_A^2}{F_{\pi}^2}\,\tau_{1,z}\,
{\rm e}^{i{\bf q}\cdot {\bf R}}\, \delta({\bf r})\,I(0)
+1 \rightleftharpoons 2\ ,
\end{eqnarray}
where ${\bf R}=({\bf r}_1+{\bf r}_2)/2$ denotes the two-nucleon
center-of-mass position, the functions $h_\pi({\bf r})$ and $\tilde{f}_\pi(r)$ are defined as
\begin{eqnarray}
h_\pi({\bf r})&=& \frac{1}{8\pi} \int_{-1/2}^{1/2} {\rm d}y \, {\rm e}^{i\, y \, {\bf q}\cdot {\bf r}}
\, \frac{{\rm e}^{-L\, r}}{L} \ , \nonumber\\
\tilde{f}_\pi(r)&=&\int_{\bf p} {\rm e}^{i{\bf p}\cdot {\bf r}} \, 
\frac{1}{(p^2+m_\pi^2)^2}=\frac{1}{8\pi} \frac{{\rm e}^{-m_\pi r}}{m_\pi} \ ,
\end{eqnarray}
where
\begin{equation}
L =\sqrt{m_\pi^2+q^2\, \left(1/4-y^2\right)} \ ,
\end{equation}
the gradients (or partial derivatives) ${\bm \nabla}$, ${\bm \nabla}_1$, and ${\bm \nabla}_2$ act
on the variables ${\bf r}$, ${\bf r}_1$, and ${\bf r}_2$, respectively, and
\begin{eqnarray}
I({\bf q})&=& \int{\rm d}{\bf x}\, e^{-i{\bf q}\cdot {\bf x}}\, \Big[ C_S\, \delta_{\alpha\beta}
-C_T\big( 2\, \sigma_{1,\alpha}\, \sigma_{2,\beta}\nonumber\\
&-&{\bm \sigma}_1\cdot {\bm \sigma}_2 \, \delta_{\alpha\beta}\big) \Big] \left[\partial_\alpha f_\pi(x)\right] \left[\partial_\beta f_\pi(x)\right]\ ,
\end{eqnarray}
Regularized expressions are obtained via the replacements
\begin{eqnarray}
h_\pi({\bf r}) \rightarrow h_\Lambda({\bf r})&=&\int_{-1/2}^{1/2} {\rm d}y \, {\rm e}^{i\, y \, {\bf q}\cdot {\bf r}}
\int_{\bf p} {\rm e}^{i{\bf p}\cdot {\bf r}} \nonumber\\
&\times&\frac{C_\Lambda(p)}{(p^2+L^2)^2} \ ,\\
\tilde{f}_\pi(r) \rightarrow \tilde{f}_\Lambda(r) &=&\int_{\bf p} {\rm e}^{i{\bf p}\cdot {\bf r}} \, 
\frac{C_\Lambda(p)}{(p^2+m_\pi^2)^2} \ , \\
\delta({\bf r})\rightarrow g_\Lambda(r)&=&\int_{\bf p} {\rm e}^{i{\bf p}\cdot {\bf r}} \, 
C_\Lambda(p) \ .
\end{eqnarray}
Lastly, we observe that i) $\rho^{(1)}_{\rm e}+\rho^{(1)}_{\rm j}$
is proportional to $I({\bf q})-I(0)$, and this quantity remains finite for any
$q$ value; ii) the requirement $\rho^{(1)}=0$ at ${\bf q}=0$ is satisfied
also when the cutoff $\Lambda$ is included.
\end{document}